\documentclass{aa}
\usepackage{txfonts}
\usepackage{natbib}
\bibpunct{(}{)}{;}{a}{}{,} % to follow the A&A style
\usepackage{graphicx}
\begin{document}

%\title{A complete high resolution three dimensional extinction map of the Galactic Bulge}
\title{Mapping the Milky Way bulge at high resolution: \\ the 3D dust extinction, CO, and X factor maps} 
%\author{M. Schultheis \& B.Q. Chen et al.}

\author{M. Schultheis \inst{1}
\and   B.Q. Chen \inst{2,3}
\and   B.W. Jiang \inst{3}
\and  O.A. Gonzalez \inst{4}
\and  R. Enokiya \inst{5}
\and  Y. Fukui \inst{5}
\and  K. Torii \inst{5}
\and  M. Rejkuba \inst{6}
\and   D. Minniti \inst {7,8,9}
}

   \institute{ Universit\'e de Nice Sophia-Antipolis, CNRS, Observatoire de C\^ote d'Azur, Laboratoire Lagrange, 06304 Nice Cedex 4, France 
 e-mail: mathias.schultheis@oca.eu
 \and
 Department of Astronomy, Peking University, Beijing, P. R. China, e-mail:bchen@pku.edu.cn
  \and
  Department of Astronomy, Beijing Normal University,
    Beijing 100875, P.R.China, e-mail: bjiang@bnu.edu.cn
    \and
    European Southern Observatory, Alonso de Cordova 3107, Vitacura, Santiago, Chile, e-mail: ogonzale@eso.org
	\and
  Department of Physics, Nagoya University, Furo-cho, Chikusa-ku, Nagoya, Aichi, 464-8601, Japan
    \and
    European Southern Observatory, Karl-Schwarzschild-Strasse 2, D-85748 Garching, Germany, e-mail: mrejkuba@eso.org
    \and
  Departamento Astronom\'iay Astrof\'isica,  Pontificia Universidad Cat\'olica de Chile, Av. Vicu\~na Mackenna 4860, Stgo., Chile, e-mail: dante@astro.puc.cl
           \and
Vatican Observatory, V00120 Vatican City State, Italy
  \and 
Departamento de Ciencia Fisicas, Universidad Andres Bello, Santiago, Chile 
 }

\abstract {Three dimensional  interstellar extinction maps provide a powerful tool for stellar population analysis. However, until now, these 3D maps were rather limited by  sensitivity and  spatial resolution.}
 {We use data from the VISTA Variables in the Via Lactea survey  together with the Besan\c{c}on stellar population synthesis model of the Galaxy to determine interstellar extinction as a function of distance in the  Galactic bulge covering $-10^{\circ}< l < 10^{\circ}$ and $-10^{\circ}< b <5^{\circ}$.}
{We   adopted a recently developed method  to calculate the colour excess. First we  constructed the H--Ks vs. Ks and J--Ks vs. Ks colour-magnitude diagrams  based on the  VVV catalogues that matched  2MASS. Then,  based on the temperature-colour relation for M giants and the  distance-colour relations, we derived the extinction as a function of distance. The observed colours were shifted to match the intrinsic colours in the Besan\c{c}on model as a function of distance iteratively. This created an extinction map with three dimensions: two spatial and one distance dimension along each line of sight towards the bulge.}
{We present  a 3D extinction map that covers the whole VVV area with a  resolution of 6\arcmin $\times$ 6\arcmin\, for J--Ks and H--Ks using distance bins of 1\,kpc. The high resolution and depth of the photometry allows us to derive extinction maps for a range of distances up to 10 kpc  and up to 30 magnitudes of extinction in $\rm A_V$ (3.0\,mag in $\rm A_{Ks}$). Integrated maps show the same dust features and consistent values as other 2D maps. We discuss the spatial distribution of dust features in the line of sight, which suggests that there is much material in front of the Galactic bar, specifically between 5-7 kpc.  We compare our dust extinction map with the high-resolution $\rm ^{12}CO$ maps (NANTEN2) towards the Galactic bulge, where we find  a good correlation between $\rm ^{12}CO$ and $\rm A_{V}$. We determine  the X factor  by combining the CO map and  our dust extinction map. Our derived average value  X=$\rm 2.5 \pm 0.47 \times 10 ^{20}cm^{-2}K^{-1}km^{-1}s$  is consistent with the canonical value of the Milky Way. The X-factor decreases with increasing extinction.}
 {}


\maketitle

\titlerunning{3D extinction map}
\authorrunning{Chen et al.}

\section{Introduction}

To study the 3D structure of the Galactic bulge one needs to know the 3D extinction correction. For example, the VISTA Variables in the Via Lactea (VVV) survey (\citealt{minniti2010}) aims to obtain the 3D  distribution of old RR Lyrae stars in the bulge.
In the large-scale studies of the bulge giants, most of the authors do not know the distance to the individual objects and therefore assume that all the stars are at the same distance, they then use 2D maps  to correct for extinction. We show here that these assumptions cause large errors.

The Galactic bulge is a main component of our Galaxy  that contains a large amount of interstellar gas and dust.  While there are a few low-extinction fields (such as Baade's window), most of the bulge is affected by high and clumpy interstellar absorption (see e.g. \citealt{schultheis1999}; \citealt{gonzalez2012}) that reaches extreme values of $\rm A_{V} > 50^{m}$ (\citealt{schultheis2009}) close to the Galactic centre (GC). An exact knowledge of interstellar extinction is crucial  to understand the different stellar  populations in the Galactic bulge.

%To study the Bulge, the extinction is a very important information. The interstellar extinction in the Galaxy Bulge is very complex, where there is very high extinction (Galactic Centre) and some place very low extinction (Baade window). 

During the past decade several two dimensional interstellar dust extinction maps have been  made for the  Galactic bulge. \citet{schultheis1999} and  \citet{dutra2003}  presented  for the first time high resolution (2' $\times$ 2') maps in the inner bulge ($ \rm -2^{\circ} < b < 2^{\circ}$ ) using DENIS and 2MASS data.   \citet{nidever2012} determined high-resolution $\rm A_{Ks}$ maps using GLIMPSE-I, GLIMPSE-II, and GLIMPSE-3d data based on the  Rayleigh-Jeans colour-excess (RJCE) method. Based on  the H--[4.5] colour, they mapped
the interstellar absorption  along  the  Galactic plane ($\rm |b| < 2^{o}$) with a spatial resolution of $\rm 2'$.

 The VVV survey  covers about 315 sq.deg of the Galactic bulge ($-10^{\circ}<l < 10^{\circ}$ and $-10^{\circ}<b<5^{\circ}$). \citet{gonzalez2012} presented for the first time the complete extinction map of the Bulge. They measured the mean (J--Ks) colour of the red clump giants in small subfields  and compared them with  a  reference field (Baade's window) to obtain reddening values E(J--Ks) for the whole Bulge area with a resolution between 2\arcmin and 6\arcmin. This map extends beyond the inner Milky Way Bulge  to higher Galactic latitudes and is the most complete study extending up to $\rm A_{V} \sim 35^{m}$, which are much higher extinction values than were determined in most previous studies. This high-resolution map allowed \citet{gonzalez2012} to demonstrate $\rm A_{Ks}$ variation of up to 0.1\,mag within a 30' field centred on  Baade's Window. In more reddened inner regions of the Bulge the extinction variation is much stronger  over smaller spatial scales as well.

In addition to near-infrared,  the optical surveys OGLE and MACHO  (\citealt{sumi2004}; \citealt{kunder2008}; \citealt{nataf2013}) provided highly accurate extinction maps of Bulge fields located mainly in the  intermediate and outer Bulge ($ \rm b > 2^{\circ}$). However, the disadvantage of optical extinction maps is the strong variation of the extinction coefficient for different lines of sight (see e.g.  \citealt{udalski2003}; \citealt{nidever2012})

All these studies  are 2D-maps that assume that all stars are located at a distance of $\rm \sim 8\,kpc$ from the Sun with no distance information. Stellar population synthesis models of the Milky Way can be used
to infer the three-dimensional distribution of interstellar extinction. \citet{marshall2006} presented  a 3D  extinction model of the Galaxy by using the 2MASS data and the stellar population synthesis model of the Galaxy, the so-called  Besan\c{c}on model of the Galaxy \citep{robin2003}. Assuming a distance--colour relation (e.g., J--Ks), they compared the observed colours for each line of sight  with the synthetic ones and attributed the corresponding distances derived from the  model.

 However, the sudy of \citet{marshall2006} is limited by the confusion limit of 2MASS in the Galactic bulge region and by the limiting sensitivity of 2MASS in highly extincted regions ($\rm A_{V} > 30^{m}$). Using a  similar  method, \citet{chen2012} used the GLIMPSE data together with the VVV data, combining the improved  Besan\c{c}on model (\citealt{robin2012}) to present colour-excess maps for six different combination of colours. Their maps agree  excellently well  with the red-clump extinction map of Gonzalez et al. (2012). The maps of  Chen  et al. are restricted to the inner Bulge ($-2^{\circ}<b<2^{\circ}$), and both \citet{marshall2006} and \citet{chen2012} only have  a spatial resolution of 15'x 15', which is too coarse for most Bulge studies.

In this  paper  we  use  the bulge  VVV data set that covers about  $\sim$315 sq. degrees to trace the 3D extinction at a high resolution  of 6'x 6'. This resolution enables for the first time  dereddening of  stars in 3D studies of the entire Galactic bulge.

%  The structure of the Galactic Bulge is believed to be dominated by the Bar (Stanek et al. 1994). Recent analysis of the RC stars across the bulge area suggested a more complex X-shaped structure (McWilliam et al. 2010; Zoccali 2010; Nataf et al. 2010; McWilliam \& Zoccali 2010; \citet{saito2011}). \citet{gonzalez2012} also used the RC stars in the VVV to trace the position of the Galactic bar, while the RC stars only shows part of the structure of the X-shape morphology. The 3D extinction can also inflect the bar of the Bulge. \citet{marshall2006} tried to fits that minimises the mean absolute deviation of the points of the highest density along the Galactici bar and find an angle of $\phi = 30 \pm 5^{\circ} $.% But because of the low resolution and the completeness limit, they can't get a more detail structure. Using the high resolution 3D extinction maps we traced the dust bar of the Galaxy and find also a X-shape structure. 

Furthermore, we  use this information to evaluate the so-called X-factor, which relates the amount of CO and molecular hydrogen defined as

\begin{equation}
 X=\frac{N_{H_2}}{W } \hspace{1cm} [\rm  cm^{-2}K^{-1}km^{-1}s],
\end{equation}

where W is the CO brightness temperature and $N_{H_2}$ is the column density of $H_{2}$.
Determining this factor empirically in different environments is important for studies of higher redshift sources where the X-factor has to be assumed, 
because it cannot be measured. Here we complement the VVV NIR dataset with the J=1-0 transition of the $^{12}CO$ map from the NANTEN2 telescope (\citealt{enokiya2013})  to investigate the relation between the extinction due to dust and the CO emission in the innermost 200 pc region of the Milky Way.

Previous work has provided evidence of  a  constant X-factor  in the Milky Way and in the Local Group with a value of  $\rm 10^{20} cm^{-2}K^{-1}km^{-1}s$ for the Galactic clouds (\citealt{solomon1987}; \citealt{young1991}; \citealt{dame2001}). \citet{glover2011} derived a relationship between the X factor and the mean extinction by analysing the global properties of the X-factor in different molecular models. They found that the X-factor  decreases with increasing mean extinction. \citet{shetty2011a} applied the radiative transfer calculation to the molecular clouds. They obtained a similar averaged X-factor for the Milky Way (about $ 2 \times 10 ^ {20} cm^{-2}K^{-1}km^{-1}s $), but found a more complex relation between the X-factor and the extinction $\rm A_V$ for different models. We here determine the X-factor and correlate the
CO gas distribution with  our 3D interstellar-dust extinction map.

We first briefly introduce  our data set and method in Sect.~2. In Sect.~3 we present the complete high-resolution 3D-extinction and compare our map with  previous work. We discuss the comparison between our extinction maps and the CO map in Sect.~5. We conclude in Sect.~6.

\section{Data, model, and method}

We uses the J, H, and Ks-band catalogues from the VVV survey. The observations cover about 315 sq.deg of  the  Galactic bulge  with $-10 < l < 10^{o}$ and $-10 < b < 5^{o}$ (\citealt{saito2012}).  Owing to saturation of the VVV data, the 2MASS catalogue  was used for $Ks < 12 \,mag$ sources, while the VVV catalogues cover $Ks > 12 \,mag$.  The  VVV data processing,  photometric catalogues construction, and a   comparison between VVV and 2MASS is presented in detail in \citet{gonzalez2011}. \citet{gonzalez2012} used these catalogues to construct the complete high spatial resolution 2D extinction map of the Bulge using the red clump star technique, while \citet{chen2012} based on the VVV and 2MASS low-resolution 3D extinction maps of the inner Bulge.
%. They are the same ones used for the present study. As it has been mentioned in \citet{chen2012}, the 2MASS data is not complete in the range $Ks < 12 \,mag$ and where replaced by the VVV data.

Our method of constructing a 3D extinction map is described in detail  in  \citet{chen2012}. We  briefly summarize it  here: We used the newly improved  Besan\c{c}on Galaxy model \citep{robin2012},  which includes a two-component bar/bulge model and applied several important corrections:
 (i)  a new  temperature-colour relation for M giants  using the  isochrones from \citet{girardi2010} and extending the grid in the $\rm T_{eff}$ vs. log\,g plane (see also \citealt{chen2012}); (ii) in contrast to  \citet{marshall2006}, we did not exclude M dwarfs from our analysis; (iii) we used  bootstrapping to derive a more realistic uncertainty in our method; (iv)the completeness limit was  calculated for each filter for each subfield   and  the observations and  the model  were cut accordingly.

\begin{table*}[htbp!]
\caption{Extinction as a function of Galactic longitude, latitude, and distance based on VVV data.}
\centering
\tabcolsep 5.8pt
\begin{tabular}{cccccccc}
\hline\hline                       
l	&	b	&$E(J-Ks)_{1-10\,kpc}$&$E(H-Ks)_{1-10\,kpc}$&$\sigma E(J-Ks)_{1-10\,kpc}$&$\sigma E(H-Ks)_{1-10\,kpc}$ \\
\hline
\end{tabular}
\tablefoot{For each position we provide the E(J--Ks),  E(H--Ks), and the corresponding sigma for each distance bin starting from 1 to 10\,kpc. }
\end{table*}

\begin{figure*}[!htbp]
   \includegraphics[width=18cm]{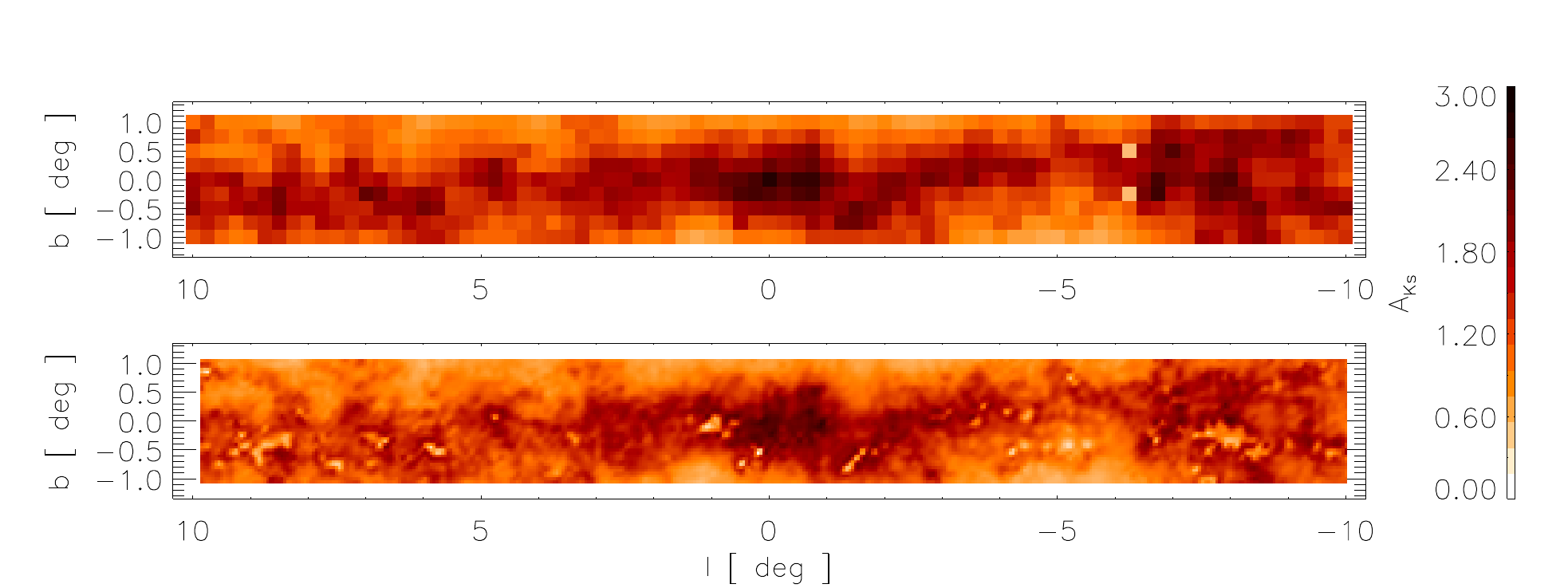}
\caption{Visual comparison of the  low-resolution extinction map (Chen et al. 2013) with our high-resolution map, integrated up to a distance of 8\,kpc.}
\label{complow-high}
\end{figure*}

\begin{figure}[!htbp]
   \includegraphics[width=9cm]{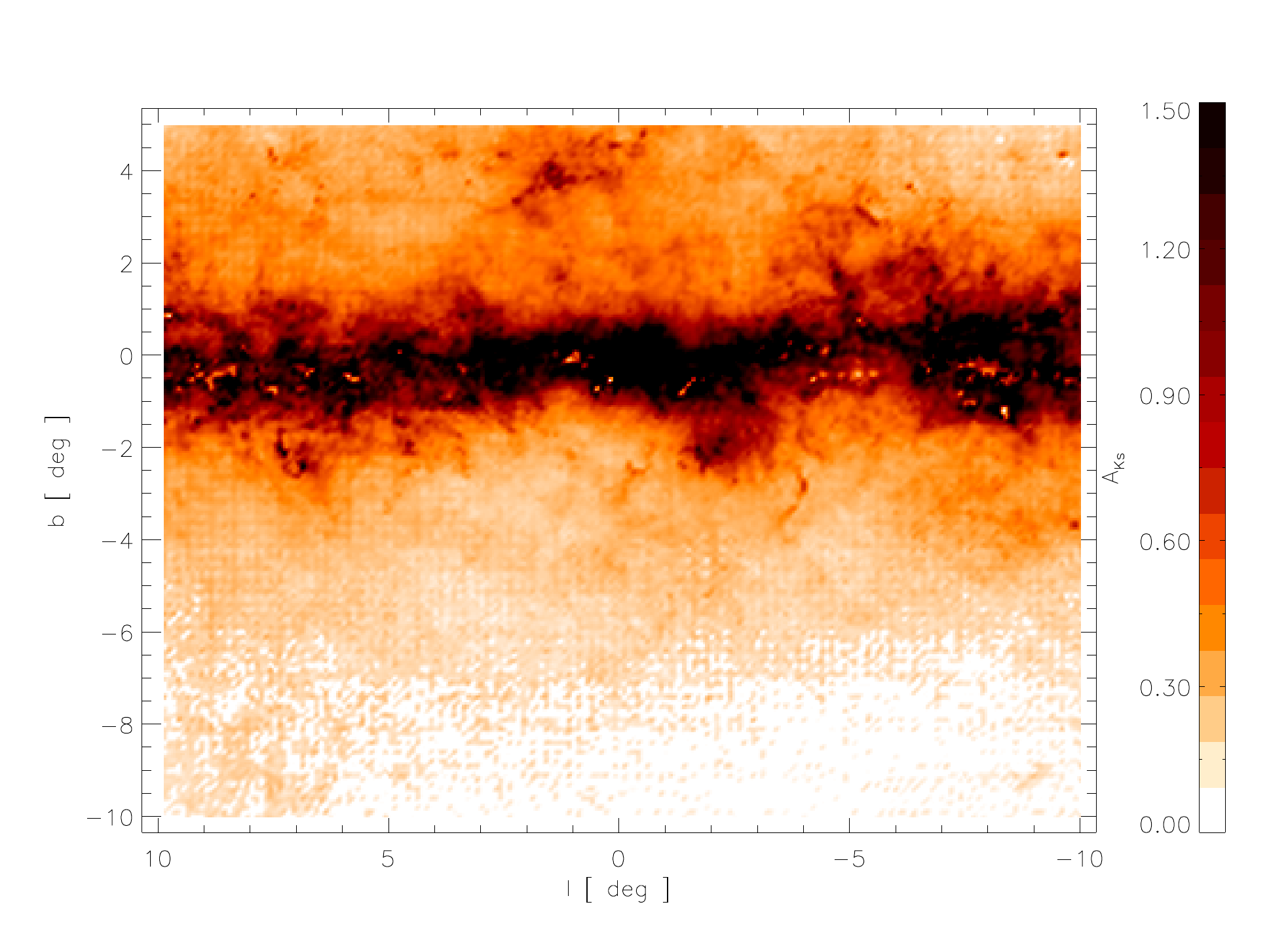}
\caption{2D extinction $\rm A_{Ks}$ integrated at 8\,kpc. The x-axis denotes the  Galactic longitude, the  y-axis the Galactic latitude. }
\label{2dextin}
\end{figure}

The data from both the VVV survey and the Besan\c{c}on model  were divided into subfields of 6' $\times$ 6' for which extinction and distances were calculated. First we added photometric errors and  diffuse extinction to the intrinsic colours $Co_{ins}$ of the model. We assumed a diffuse extinction of 0.7 mag/kpc in the V band and a constant photometric error of 0.05\,mag \citep{saito2012}. The colours were then sorted for the observed and simulated data and normalised by the corresponding number of objects in each bin given by  $n_{bin}=floor(min([N_{obs},N_{bes}])/100)$.  $N_{obs}$ and $N_{bes}$ are the total number of observed stars (in each subfield) and the stars in the Besan\c{c}on model. This ensured that we had at least 100 stars per bin. In each corresponding colour bin, the $Co_{ins}$ of the stars from the model data and the $Co_{obs}$  from the observational data were used to calculate the extinction using the following equations: 

\noindent
\begin{center} 

$A_{Ks,J-Ks} = 0.528 \times (\overline{(J-Ks)_{obs}}-\overline{(J-Ks)_{ins}}) $\\
$A_{Ks,H-Ks} = 1.61 \times  (\overline{(H-Ks)_{obs}}-\overline{(H-Ks)_{ins}}). $\\

\end{center}

We assumed  the interstellar extinction law of \citet{nishiyama2009}. A first distance and extinction estimate were thus obtained and directly applied to the intrinsic magnitudes in the model, which provided a  new simulated colour. We constructed histograms for each colour of the new simulated data and the observational data  using the same bin size (0.05\,mag). The $\chi ^2$ statistics \citep[][]{press1992,marshall2006} was then used to evaluate the similarity of the two histograms, given by

\begin{figure*}[!htbp]
   \includegraphics[width=18cm]{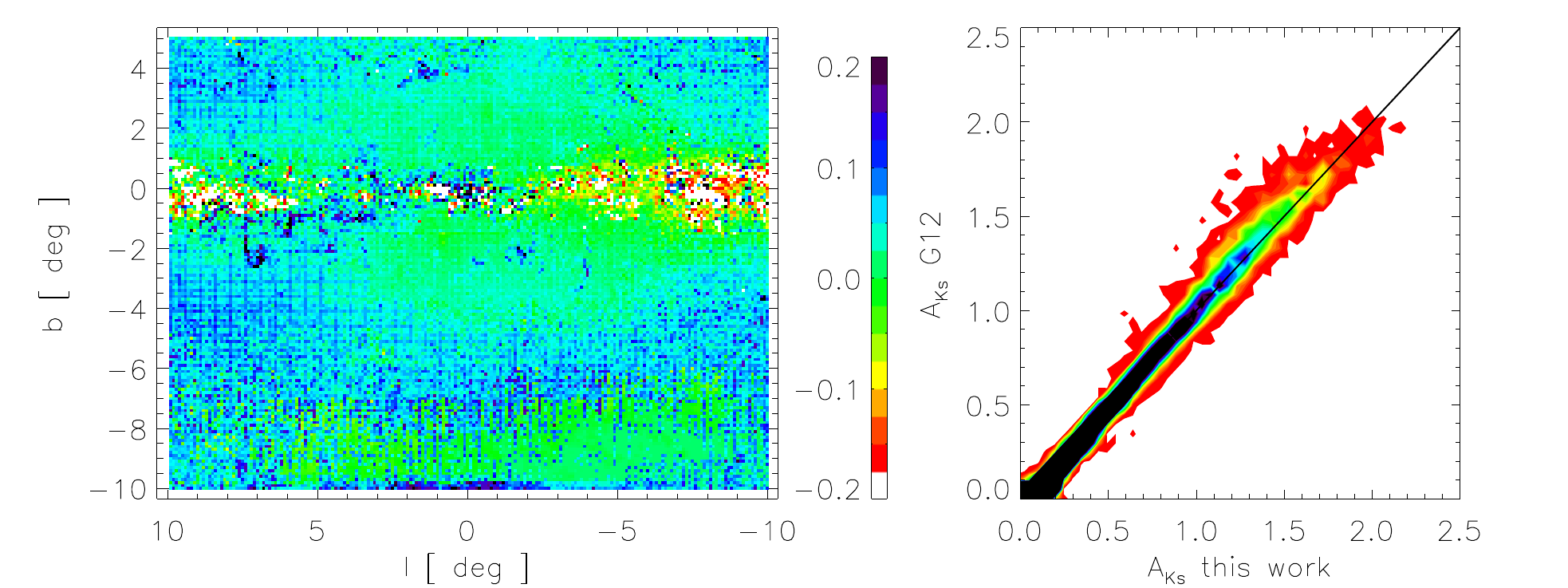}
\caption{Left panel: difference in the 2D extinction ($\rm A_{Ks_{G12}} -  A_{Ks}$) as a function of Galactic
 longitude and latitude compared with G12. Right panel: $\rm A_{Ks_{this work}}$ vs. $\rm A_{Ks_{G12}}$}
\label{2dosc}
\end{figure*}

\begin{equation}
 \chi^2 = \sum _j {(\sqrt{N_{obs}/N_{sim}}n_{sim_j}-\sqrt{N_{sim}/N_{obs}}n_{obs_j})^2/(n_{sim_j}+n_{obs_j})},
\end{equation}
where $N_{obs}$ and $N_{sim}$ are the total number of observed and simulated stars in each subfield, while $n_{obs_j}$  and $n_{sim_j}$  show the number of stars for the $j$th  colour bin  of the observations and simulated data. The first extinction estimate applied provided a new set of simulated data. In this new data set, some of the stars may be fainter and fall outside the completeness limit, while others may be brighter and now are inside the limit. The new simulated data were used iteratively to refine our results. In total, we performed 20 iterations for each subfield and each colour. The distance and extinction with the lowest $\chi ^2$ were our final result. The corresponding errors in the extinction and  distances were calculated using the bootstrap method (see \citealt{chen2012} for further details).

 The distance resolution depends on the chosen spatial resolution and the smallest number of stars, which is a function of  the stellar density and the interstellar extinction. We set the smallest number of stars in the highest extincted region to be 100 to ensure that a  proper fit to the Besan\c{c}on model can be still achieved. In less extincted regions the number density will increase, giving a larger number of stars to fit the model. To resolve small-scale extinction features together with a decent distance resolution, our distance intervals were interpolated in bins of 500\,pc with a spatial resolution of
6\arcmin $\times$ 6\arcmin.

\section{High-resolution three-dimensional extinction maps}

The complete high-resolution 3D extinction map for the region of the Bulge covered by the VVV survey  will be available in electronic form at the CDS\footnote{Table 1 is only available in electronic format from the CDS via anonymous ftp to cdsarc.u-strasbg.fr (130.79.128.5) or via http://cdsweb.u-strasbg.fr/cgi-bin/qcat?J/A+A/.}.

Table 1 provides an example of the format. Each row of Table 1 contains the information for one line of sight: Galactic coordinates along with the measured quantities for each distance bin E(J--Ks), E(H--Ks), and the respective uncertainties. These results will also be available via the BEAM calculator\footnote{http://mill.astro.puc.cl/BEAM/calculator.php}  webpage (\citealt{gonzalez2012}; \citealt{chen2012}). Users of the BEAM calculator can choose to retrieve the extinction calculation with a specific reddening law and distance interval.   

\subsection{Integrated map at 8\,kpc: the Galactic bulge extinction map}

The  three dimensional extinction maps allowed us to quantify the total amount of extinction that affects a source at a given distance, which  results from the interstellar material distributed along the line of sight.

 We  first integrated our 3D extinction map up to a distance of 8\,kpc  to produce a 3D map in units of $\rm A_{Ks}$. The projected map can be used to compare our results with the previously published 3D extinction maps from the literature. 

As described in the previous section, we increased both the resolution and the coverage of the extinction map presented in Chen et al. (2013) and obtained from this the complete 3D extinction map for the VVV region of the Bulge. Figure \ref{complow-high} shows  the comparison between the low-resolution \citep{chen2012} and high-resolution maps, both integrated up to a distance of 8\,kpc. In general, we see very similar dust features. However, the fine filamentary structure of the dust extinction can only be resolved by the high-resolution map.The extinction, in the Galactic centre  is so high that the VVV data does not contain enough stars for a small number of subfields ($\sim$ 10) when the high-resolution binning is used. These subfields were then filled-in the final high-resolution map by interpolating the lower resolution results \citep{chen2012}. The complete extinction map of the Bulge, obtained by integrating our 3D map up to 8\,kpc, is shown in Fig.~\ref{2dextin}.

\begin{figure}[!htbp]
   \includegraphics[width=7.0cm]{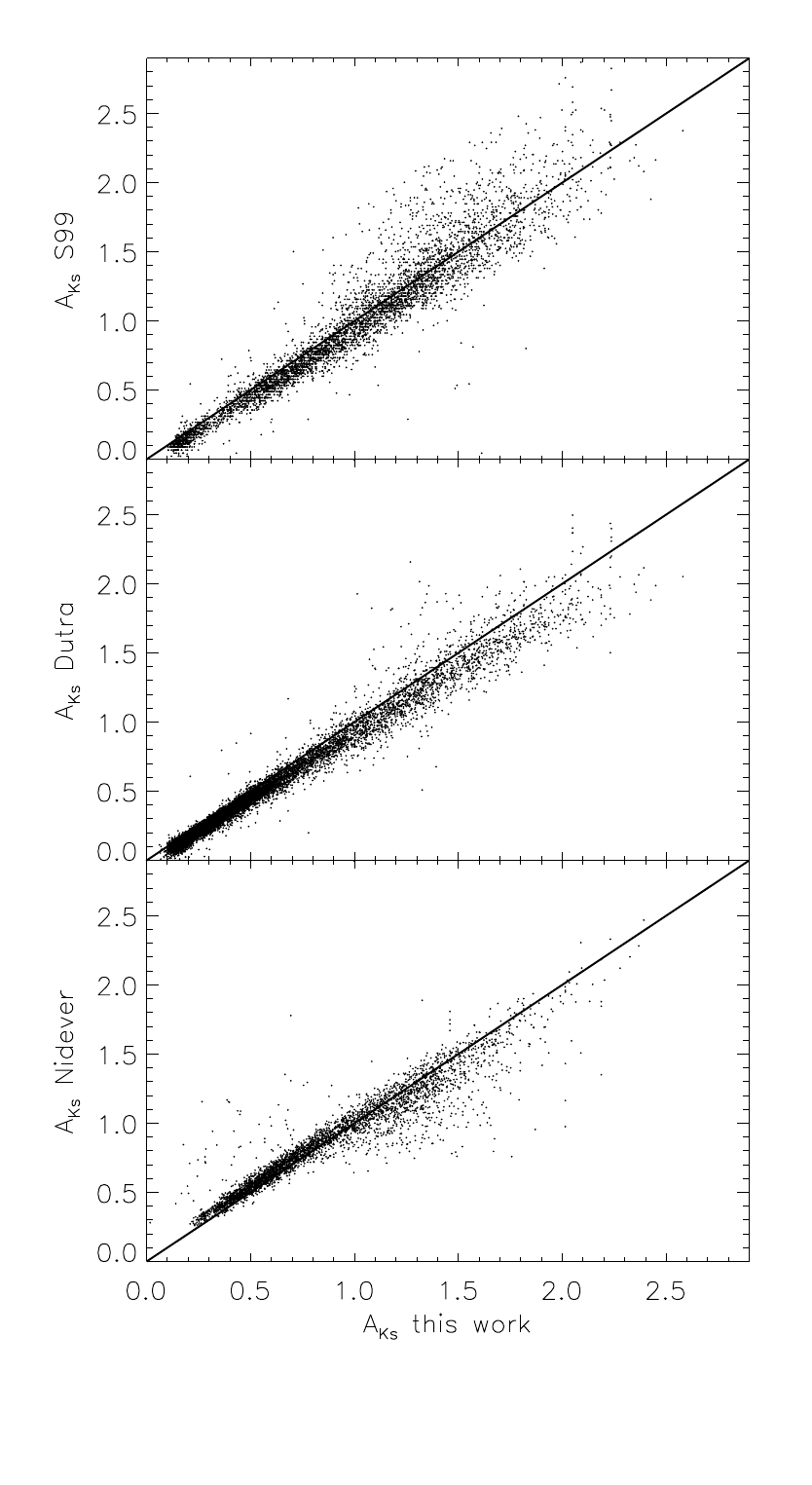}
\caption{Integrated 2D extinction  at 8\,kpc of our work  compared with Schultheis et al. (1999, top panel), Dutra et al. (2003, middle panel) and Nidever et al. (2012, bottom panel). All maps have been transformed assuming the extinction law of Nishiyama et al. ($\rm A_{Ks} = 0.528 \times E(J-Ks)$).}
\label{rjcedut}
\end{figure}

\citet{gonzalez2012} provided a complete high-resolution map (2\arcmin--6\arcmin) using the same data set,  but they only used red clump stars as tracers of extinction. They selected  the Bulge red clump stars for each line of sight from a CMD digram for each of the subfields and determined the interstellar extinction.

\begin{figure*}[!htbp]
   \includegraphics[width=18cm]{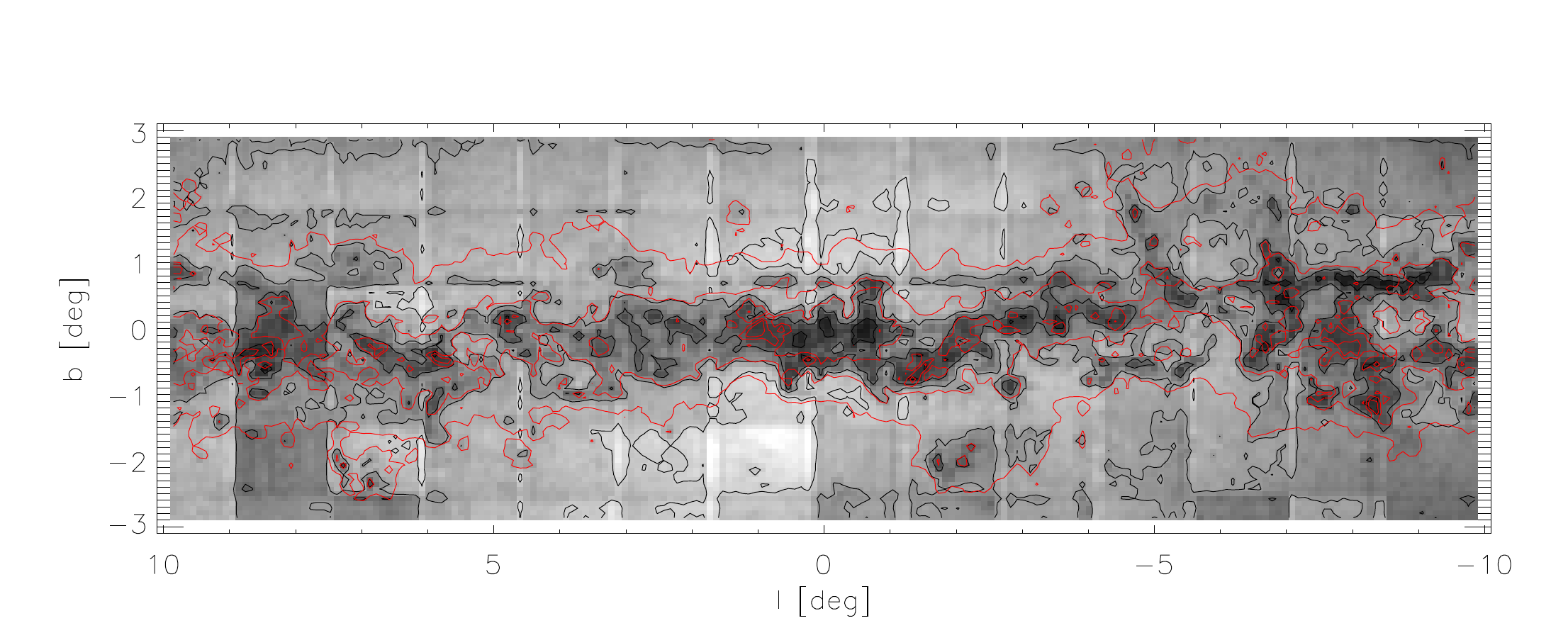}
\caption{Greyscale of the star count distribution together with the contour of the 2D extinction. The lower star density is denoted in black and higher star density in white. Note that  stars were counted twice in the overlapping regions of the different tiles (seen as white vertical stripes).}
\label{aksc}
\end{figure*}

The red clump stars are a very homogeneous population where effects such as metallicity play only a minor role, and they are predominantly located in the bulge at approximately the same distance, which makes them excellent tracers for the 2D extinction map. The difference between the mean (J--Ks) colour of the selected RC stars for each subfield with respect to that of  Baade's Window, where the extinction is well known, was used to compute the average colour excess per subfield. We converted the $E(J-Ks)$ colour excesses of \citet{gonzalez2012}  into $A_{Ks}$ using the Nishiyama extinction law ($A_{Ks}=0.528 \times E(J-Ks)$)  to compare it  with our extinction map.

The left panel of Fig.~\ref{2dosc} shows the difference $\rm \Delta A_{Ks}$  between the map of  Gonzalez et al. (G12) and our map as a function of Galactic longitude and latitude, the right panel shows $\rm A_{Ks}$ from our work vs. $\rm A_{Ks_{G12}}$. Although completely different methods were used to produce the maps,  the two methods agree quite well, with a  mean difference lower than 0.1\,mag in $A_{Ks}$. This comparison shows that  our maps are reliable.
 There is a larger scatter for highly extincted regions ($\rm A_{Ks} > 1.5$) located close to the
 Galactic plane. The number densities decreases quite significantly there (see \citealt{chen2012})  and  larger photometric uncertainties (due to the fainter Ks magnitudes) result in   larger errors in $A_{Ks}$.  The G12 map predicts higher extinction there,  but as pointed out by \citet{gonzalez2012}, the RC stars in these very high extinction regions suffer from severe incompleteness.

\begin{figure*}[!htbp]
   \includegraphics[width=18cm]{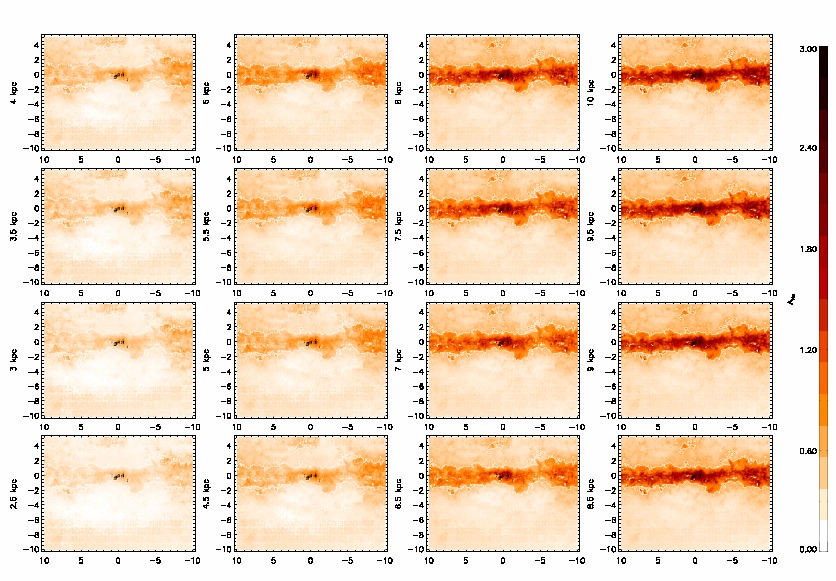}
\caption{Integrated 3D extinction $\rm A_{Ks}$. The x-axis denotes the Galactic longitude, the  y-axis the Galactic latitude. The units are $\rm \delta A_{Ks}\,kpc^{-1}$. The distances are given with respect to the sun. White contours mark
the highest extincted regions with $\rm A_{Ks} \geq 0.8$. The tilt in the dust extinction with the negative longitude part located above the plane and the positive longitude below the plane is clearly seen. The indicated feature located at $\rm l \sim 2^{\circ}$, $\rm b \sim 4^{\circ}$ is the Pipe nebula.}
\label{3dextin}

\end{figure*}
\begin{figure*}[!htbp]
   \includegraphics[width=18cm]{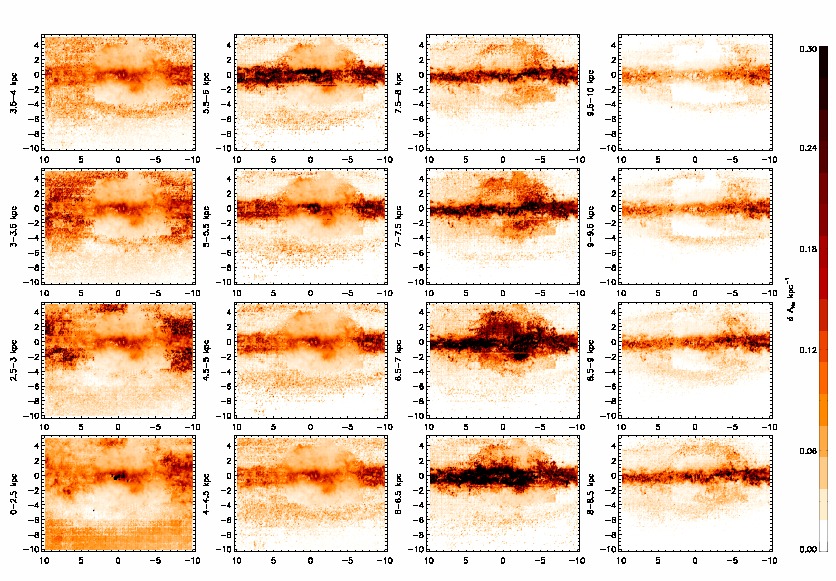}
\caption{Differential 3D extinction $\rm A_{Ks}$. The x-axis denotes the Galactic longitude, the  y-axis the Galactic latitude. The units are $\rm \delta A_{Ks}\,kpc^{-1}$. The distances are given with respect to the sun.}
\label{3dextindiff}
\end{figure*}

%\begin{figure*}[!htbp]
%\includegraphics[width=17cm]{3dextinvvv15_jk.pdf}\llap{\raisebox{6cm}
% {\includegraphics[width=17cm]{3dextinvvv15_diff.pdf}}}
%\caption{``Face on view'' of the integrated dust extinction (lower panel) and
 %the differential dust extinction (upper panel). The sun position is marked
% as ``X'' and the GC as ``+''. Indicated on the y-axis is the galactic longitude.}
%\label{faceon}
%\end{figure*}
Fig.~\ref{rjcedut} shows an additional  comparison of the 2D extinction maps of \citet{schultheis1999}, \citet{dutra2003}, and \citet{nidever2012}. \citet{schultheis1999} presented extinction maps for the inner Bulge with a  very high spatial resolution (2'$\times$2'). They used the DENIS near-infrared data set in combination with theoretical red giant branch (RGB)/asymptotic giant branch (AGB) isochrones from \citet{bertelli1994}. They showed that  the observed sequence matched the isochrone with suitable reddening well for an appropriate sampling area.  The highest extinction that could be reliably derived from the $\rm J$ and $\rm Ks$ data available from DENIS was about $\rm A_V=25^{m}$. \citet{schultheis1999} found that in some areas, presumably those with the highest extinctions, no J-band counterparts for sources were detected in $\rm Ks$.   For these regions, only lower limits to the extinctions could be obtained. \citet{dutra2003} obtained similar results using the same technique with 2MASS data. They presented the extinction map within $\rm 10^{\circ}$ of the Galactic centre with a resolution of 4'$\times$4'. 

 \citet{nidever2012} determined high-resolution $\rm A_{Ks}$ maps (2') using GLIMPSE-I, GLIMPSE-II, and GLIMPSE-3d data based on the  Rayleigh-Jeans colour-excess (RJCE) method using the H--[4.5] colour, which is robust against spectral type.  They obtained several extinction maps using RGB  stars, RC stars, or their complete sample of stars. We used their ``ALL'' star catalogues. All  maps were transformed using the \citet{nishiyama2009} extinction law. All three maps show   general  agreement with our map. We note  several differences: (i) \citet{dutra2003} derived systematically lower extinction values; (ii) \citet{schultheis1999} computed systematically larger  extinction for highly extincted regions, which might be mainly due to the completeness limit of DENIS; and  (iii) the RJCE method in general shows the lowest  dispersion and agrees even for the highest extincted regions.

Figure~\ref{aksc} shows the number of stars in grey scale (the lower star density is denoted in black, the  higher density in white) together with the contours of our extinction map. The number of stars and extinction are highly correlated, i.e.: the number density is smaller when the  extinction is higher. We also also
a structure at $\rm |l| < 1$ and  $\rm |b| < 1$ with high extinction and small number densities, which
 might be related to the nuclear bar (see also e.g. \citealt{robin2012}; \citealt{gonzalez2011b}; \citealt{rodriguez2008}). However, the existence and the nature of this structure is still debated (see e.g. \citealt{gerhard2012}). The white vertical stripes in Fig.~\ref{aksc} are due to double counting of sources in the overlapping regions of the adjacent VVV tiles. We did not correct for these double sources here because it has no implication on the final results.

\subsection{Integrated extinction map at different distances}

2D maps for the Galactic bulge such as the one presented in \citet{gonzalez2012} that are also based on the VVV data provide a measurement of integrated extinction at a fixed mean distance of $\sim8$\,kpc, which is the mean distance of the Bulge stars. These 2D maps  very likley overestimate the extinction for sources located between us and the Galactic bulge.

As an example exercise, we consider two globular clusters in the direction of the Galactic bulge, NGC6626 and NGC6658, which have a distance modulus of 13.60 (5.2\,kpc) and 14.30 (7.2\,kpc) magnitudes, respectively (\citealt{chun2010}). From the 3D extinction map we derive $\rm A_{K}=0.101$ for NGC6626 and $\rm A_{K}=1.294$ for NGC6658. If we were to take the extinction from the 2D map integrated to 8-kpc (which is typically assumed), the adopted reddening would be $\rm A_{K}=0.182$ and $\rm A_{K}=1.792$, resulting in an overestimation of approximately 0.7 and 4 magnitudes in $\rm A_{V}$, respectively. Photometric studies of ages and metallicities from colour-magnitude diagrams would definetly be affected if the 2D map extinction value were used instead of the the 3D extinction map  for these sources along the line of sight.

Figure~\ref{3dextin} shows examples of our integrated extinction map, projected at different distances from the Sun. The fine details of the dust 
structures at different distances are clearly visible in these high-resolution maps. The apparent tilt in the highest amount of dust extinction (indicated as white contours in Fig.~\ref{3dextin})  is also visible,  with the negative longitude (far-side) part  located above the plane and the positive longitude (near-side) part below the plane. This is probably due to the existence of the dust lanes (see \citealt{marshall2008}). Additionally, because the  sensitivity of VVV is higher than that of 2MASS, we are also able to trace the dust lanes at $\rm |l| < 2^{o}$.  Another feature seen in the maps is the Pipe nebula, which is located at about $\rm l \sim 2^{\circ}$, $\rm b \sim 4^{\circ}$, as part
of the Ophiuchus dark cloud complex.

%  We notice clearly from the high-resolution 3D extinction maps the fine details of the dust extinction  structures at different distances. The variations of the small scale extinction close to the galactic plane are visible at larger distances (beyond 7\,kpc), while the local structures at closer distances can be seen at higher galactic latitudes. Figure~\ref{3dextin} shows that the extinction at high latitude regions is significant and variable at small distances and become smoother and more homogeneous  at larger distances. In contrast, the extinction at low latitudes (the disk) does not change a lot.  Clearly visible is also the tilt in the dust extinction (indicated
%as white contours in Fig.~\ref{3dextin})   with the negative longitude (far-side) part located above the plane and the positive longitude (near-side) part below the plane. This is probably due to the existence of the dust lanes (see \citealt{marshall2008}). Due to the higher sensitivity of VVV compared to 2MASS we are also able to trace the dust lanes at $\rm |l| < 2^{o}$.}}

\subsection{Differential extinction maps and the spatial distribution of dust features}

Another application for our 3D map is that one can investigate the actual spatial distribution of the dust that is responsible for the stellar extinction of background sources. This can be investigated from our maps, as shown in Fig.~\ref{3dextindiff}, from the difference between the integrated extinctions of two subsequent distance bins  $\rm A_ {Ks,(d+0.5)kpc} - A_{Ks,(d)kpc}$ (e.g. $\rm A_{Ks,5 kpc} - A_{Ks,4.5 kpc}$).

\begin{figure}[!htbp]
\includegraphics[width=9cm]{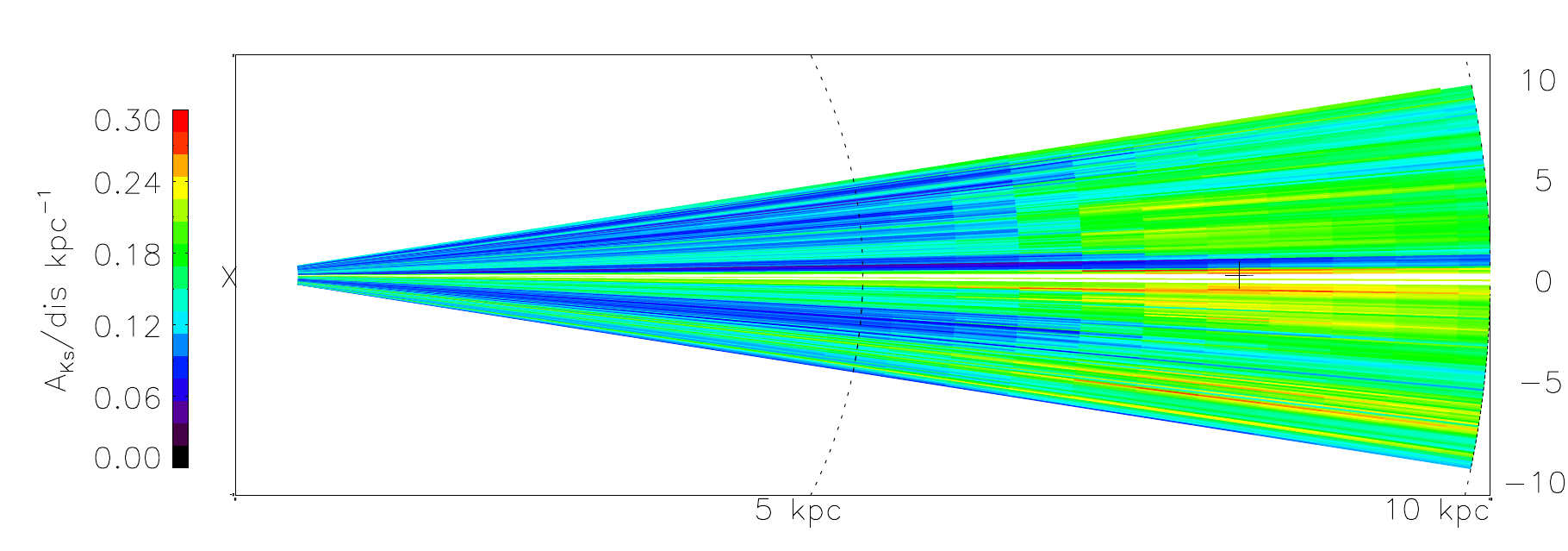}\llap{\raisebox{3.5cm}
 {\includegraphics[width=9cm]{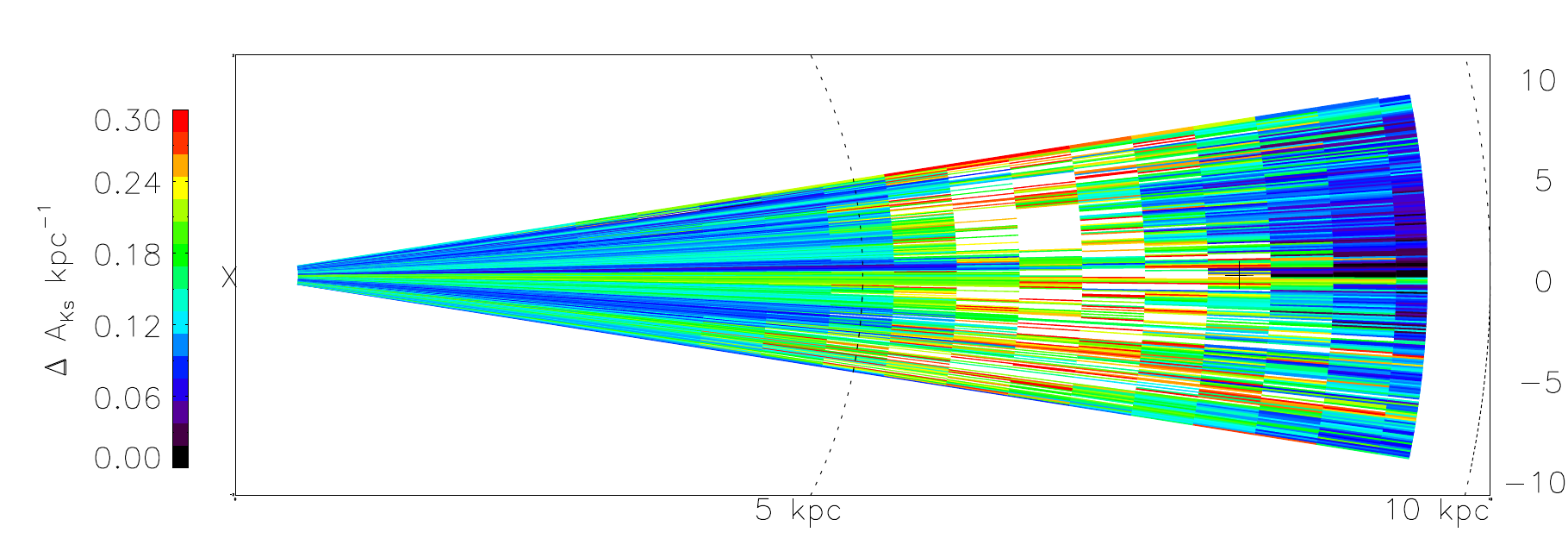}}}
\caption{Face-on view of the integrated dust extinction (lower panel) and
the differential dust extinction (upper panel). The position of the sun is marked X and that of the Galactic centre  +. The Galactic longitude is indicated on the y-axis.}
\label{faceon}
\end{figure}

Figure~\ref{faceon} shows another view of the extinction map, both an integrated and differential one, as seen from the North Galactic pole. In contrast what was reported in  \citet{marshall2006}, we see a much smoother increase in extinction for the inner 5\,kpc, while our extinction values increase beyond 5\,kpc. We confirm the elongated structure, which is much more pronounced than in \citet{marshall2006}, and  passes through the Galactic centre that is related to the dust bar.

\begin{figure*}[!htbp]
   \includegraphics[width=18cm]{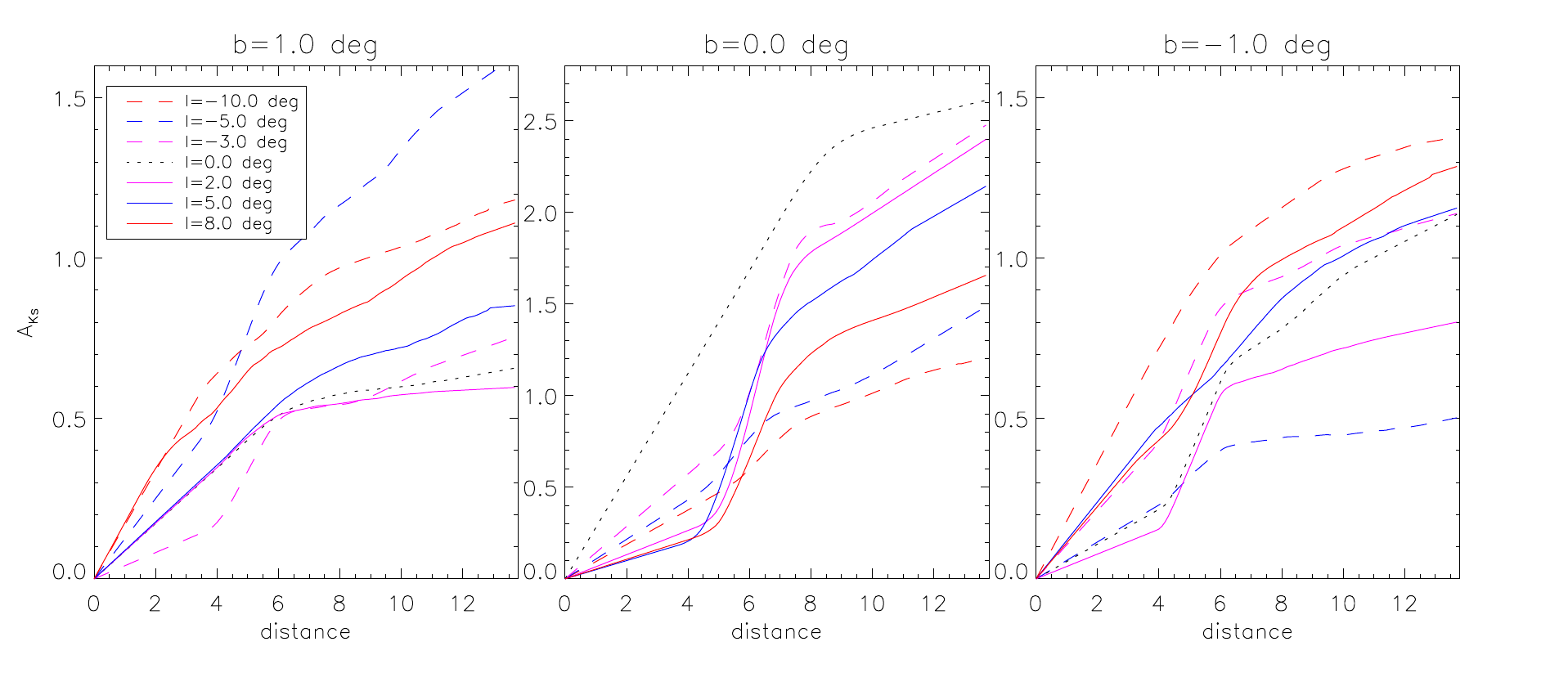}
\caption{Distance vs. $\rm A_{Ks}$ along different lines of sight. Different colours indicate different Galactic longitudes.}
\label{disak}
\end{figure*}

Figure~\ref{disak} shows the interstellar extinction $\rm A_{Ks}$ as a function of distance for several selected subfields at different Galactic latitudes located  at b=+1$^{o}$  (left), b=0$^{o}$ (middle), and b=-1$^{o}$ (right). Different colours stand for different longitudes. As expected, the extinction increases with increasing distance.  Depending on the chosen line of sight, the extinction changes in a different way. However, for most of the subfields, the extinction grows much faster between 5 and 7\,kpc, which  indicates that dust clouds are clustered in this range of distance. The extinction grows only slowly after about 8\,kpc, where in many cases the highest dust absorption is reached (see also Fig.~\ref{3dextin}).  We also note that the  dust distribution is  non-axisymmetric along the major axis.

We clearly see that the peak in the extinction increases significantly at about 5--7\,kpc and decreases after 8\,kpc. This is at odds with the intuitive scenario of Galactic extinction, where this reaches a maximum at 8 kpc, that is, at the Galactic centre. An overdensity of material, strongly concentrated between 5--7\,kpc, would then be required  to explain the sudden increase of extinction seen at these distances in Fig.~\ref{3dextindiff}. \citet{liu2012} investigated the extinction properties towards the anticentre direction, using spectra of disk red clump giants, and discovered  extinction peaks at a Galactocentric distance of 9.5 and 12.5 kpc. They identified this feature as the dust lane of the Perseus sprial arm. This finding leads us to suggest a similar scenario, in this case in the direction of the Galactic centre, to explain the extinction peak between 5--7\,kpc observed in our maps.

\begin{figure}
   \includegraphics[width=9cm]{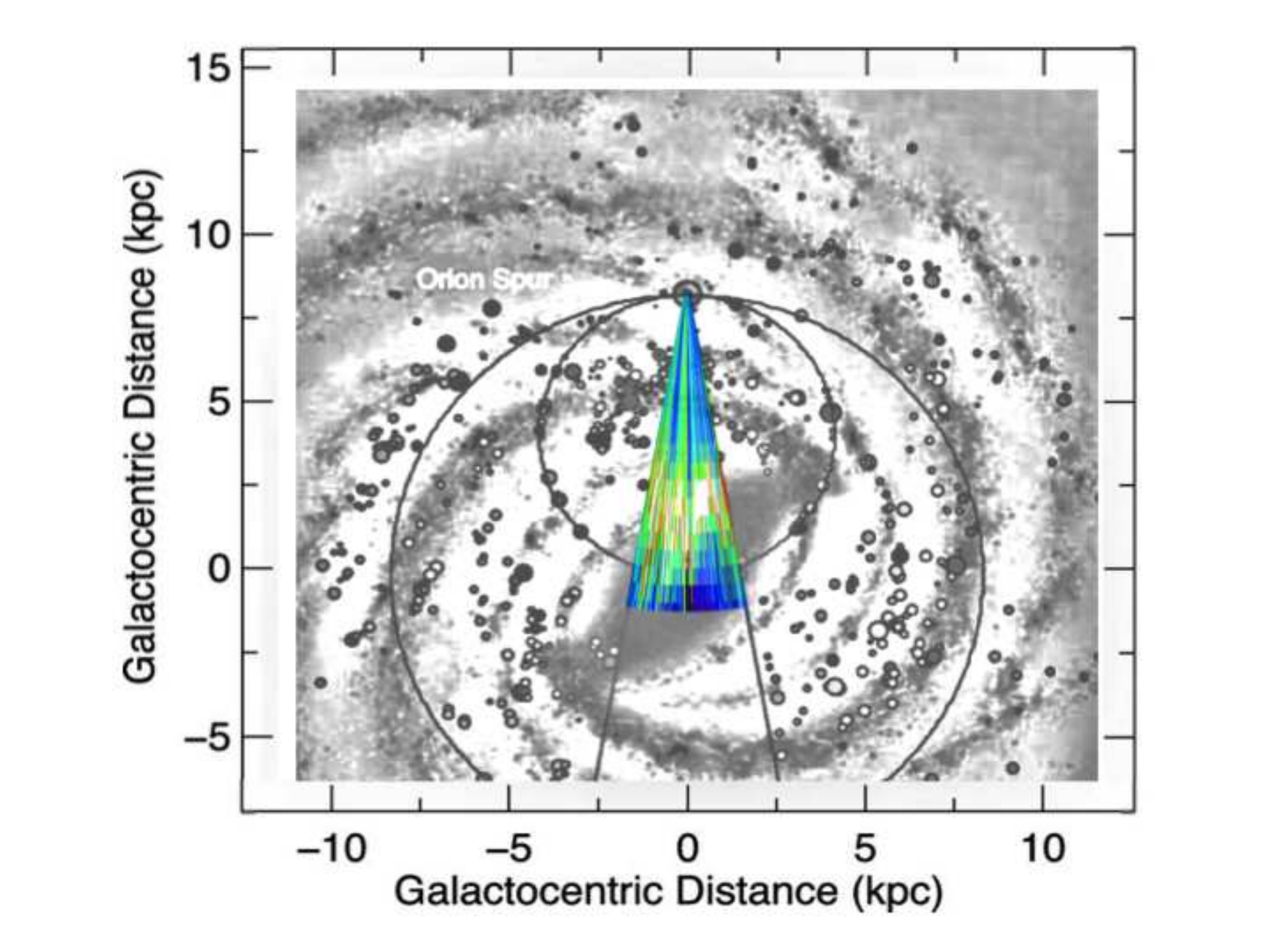}
\caption{Face-on view of the differential dust extinction shown overplotted on  the illustration of the Milky Way produced by Robert Hurt and positions of HII+ MYSOs sources detected in the Galactic disk from Fig. 6 of Urquhart et al. (2014).}
\label{faceon_disk}
\end{figure}

To test this possibility, we show in Fig.~\ref{faceon_disk} the face-on view of the differential extinction map shown in Fig.~\ref{faceon} now overploted on the illustration of the Galaxy produced by Robert Hurt of the Spitzer Science Centre that includes positions of HII+ MYSOs sources detected in the Galactic disk (adapted from Fig. 6 of \citealt{urquhart2014}). The increase of extinction observed in our map shows no clear correlation with the expected location of the closer Galactic spiral arms. However, the distance at which the most dramatic increase of extinction occurs, as seen in Figs.~\ref{3dextindiff}  Fig.~\ref{disak}, is clearly located immediately in front of the Galactic bar and appears to follow the orientation angle of the bar. This would indicate that there is a high concentration of dust material, that is, a dust lane, in front of the Galactic bar.

Alternatively, this peak of extinction at 5\,kpc might be due to the so-called molecular ring (\citealt{stecker1975}; \citealt{roman2010}), a prominent feature in the CO emission (\citealt{dame2001}) located around 4\,kpc from the Galactic centre. The existence of this molecular ring is still debated. \citet{dobbs2012} showed that  most of the CO emission in the velocity-longitude space can be fitted by nearly symmetric two-armed spiral pattern, where one of the spiral arms corresponds to the molecular ring (\citealt{dobbs2012}).

\begin{figure}[htbp]
   \includegraphics[width=9.0cm]{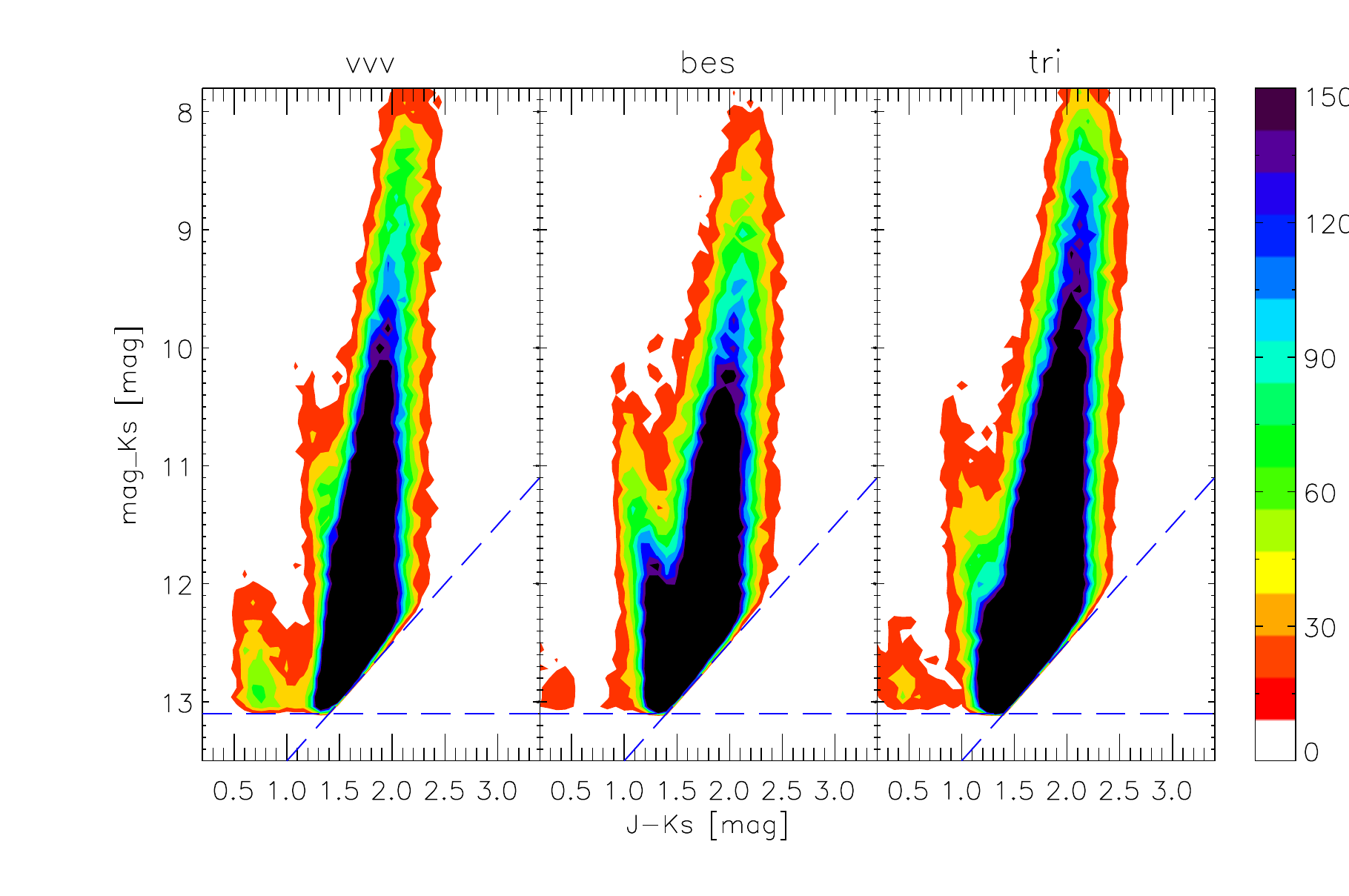}
\caption{J--Ks vs. Ks colour-magnitude diagram of the field at  $\rm l = 1.5^{o}, b = +1.5^{o}$. The left panel
 shows the VVV data, the middle panel the Besan\c{c}con model, and the right panel the TRILEGAL model. For the two models (Besan\c{c}on and Trilegal) the CMDs were broadend by adding our 3D extinction map.}
\label{comp2}
\end{figure}

\begin{figure}[!htbp]
   \includegraphics[width=9.0cm]{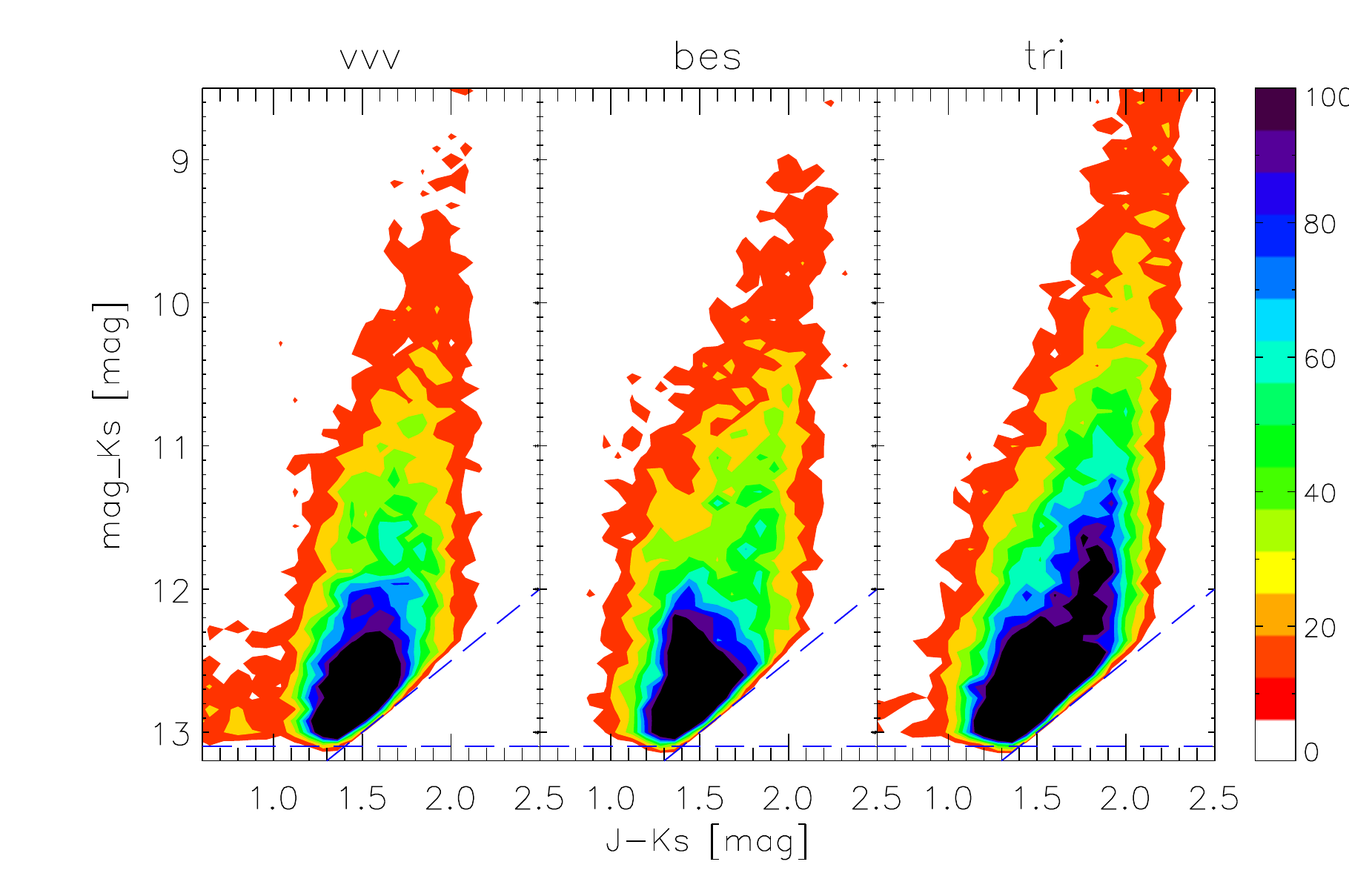}
\caption{Similar as Fig.~\ref{comp2}, but for the field $\rm l=6^{o}, b = -2^{o}$.}
\label{comp1}
\end{figure}

\section{Application to stellar population synthesis models}

Three dimensional extinction is an important ingredient for stellar population synthesis models. To compare the different stellar population synthesis models we applied our 3D extinction map also to the TRILEGAL model (Version 1.6). This allowed us to compare  these two models in more detail. The TRILEGAL model was developed by \citet{girardi2005} and is a population synthesis code for simulating the stellar photometry of any field in the Milky Way galaxy. The model has proven to  reproduce number counts well, amongst others, in all three pass-bands of the 2MASS catalogue.  The bulge component was introduced in TRILEGAL by \citet{vanhollebeke2009}, who also calibrated its stellar population and metallicity distribution using photometry from the 2MASS and OGLE-II surveys of red clump stars. \citet{uttenthaler2012} and \citet{saito2012b}   first compared  TRILEGAL and the Besan\c{c}on model and found very similar colour-magnitude diagrams. 
Figure~\ref{comp2} shows a field in the inner Bulge located at $\rm l=1.5^{o}, b = +1.5^{o}$. To trace the differences we did not normalise the model data but compared the absolute numbers. We cut the catalogues by calculating the completeness limit of VVV, indicated by the dashed line (see Fig.~\ref{comp2}). We clearly see here that our 3D extinction model matches  the observed colour-magnitude diagram well.  While  models and the VVV data the two agree well in general,  the Besan\c{c}on model predicts a too large proportion of Galactic disk K giants, which has been noted by \citet{chen2012}. It also misses  the brighter end of the Bulge sequence (K $<$ 8.5), which is visible in TRILEGAL. Figure~\ref{comp1} shows another Bulge field located at $\rm l=6^{o}, b = -2^{o}$ with slightly lower extinction. Here  the predicted colour distribution for both
Besan\c{c}on and TRILEGAL  again agree well with the VVV observed one. The colours of the TRILEGAL model 
are slightly too red and the fraction of bright Bulge giants (Ks $<$ 10) is too high. Again, the Besan\c{c}on model shows a broadening of the CMD due to the K giants in the Galactic disk. 

These two examples clearly show that the 3D extinction map allows  a  {\em{systematic comparison}} between stellar population synthesis models in general. Our derived 3D dust extinction depends  on the stellar density, which is strongly model dependent. Using the same method but applied to the TRILEGAL model  would reveal possible  systemtatic differences in the 3D dust extinction map caused by the differences in the stellar population synthesis models.

\section{Comparison with  high-resolution CO data}

Dust extinction at visible and near-infrared wavelengths is produced by large dust grains, which dominate the total dust mass of the galaxy, while the CO emission map traces the gas density and is very sensitive to the intrinsic properties of the gaseous medium, such as density, metallicity, and the background radiation field.

%The Central Molecular  Zone (CMZ) is the innermost $\sim$ 200\,pc region of the Milky Way. 
%It covers about $\rm -0.7^{o}   < l < 1.8^{o}$  in longitude  and $\rm -0.3^{o} < b < 0.2^{o}$ in latitude.  It is  a  giant molecular cloud complex with an asymmetric distribution of molecular clouds (see e.g. \citealt{morris1996}, \citealt{oka2005}).

\citet{enokiya2013} obtained observations of the J=1-0 transition of $\rm ^{12}CO$ with the NANTEN2 telescope using a high spatial resolution
of $\sim$ 200\arcsec. The observed CO intensity, I(CO), which is often expressed as an integrated brightness temperature (hereafter W), is considered to be a good tracer of the column density of molecular hydrogen $N_{H_{2}}$ using a constant X-factor.

\begin{figure}[!htbp]
   \includegraphics[width=9cm]{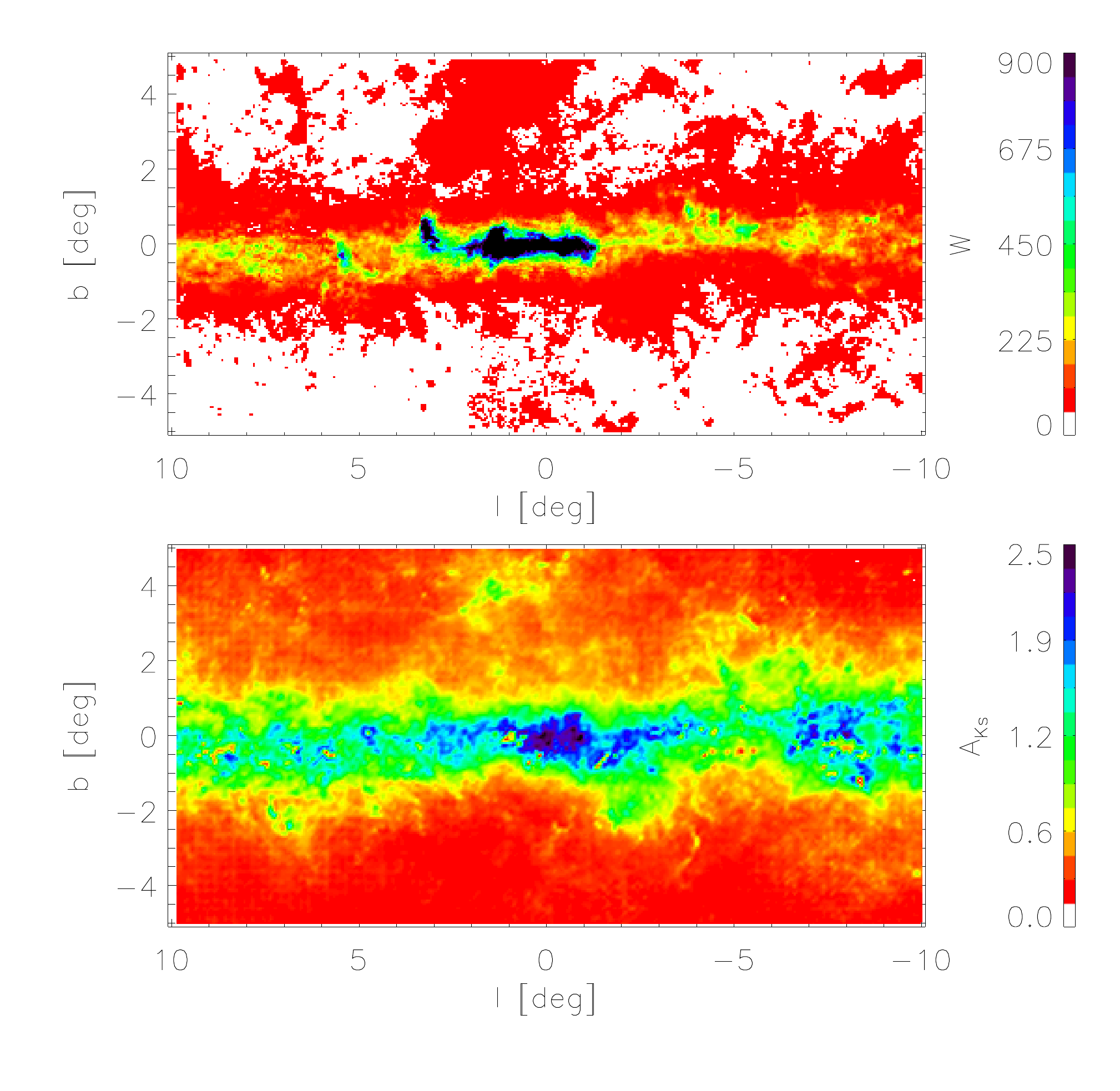}
\caption{Distribution of the CO (top) and the dust extinction map (bottom) for the inner Bulge.}
\label{akco}
\end{figure}

\begin{figure}[!htbp]
   \includegraphics[width=9cm]{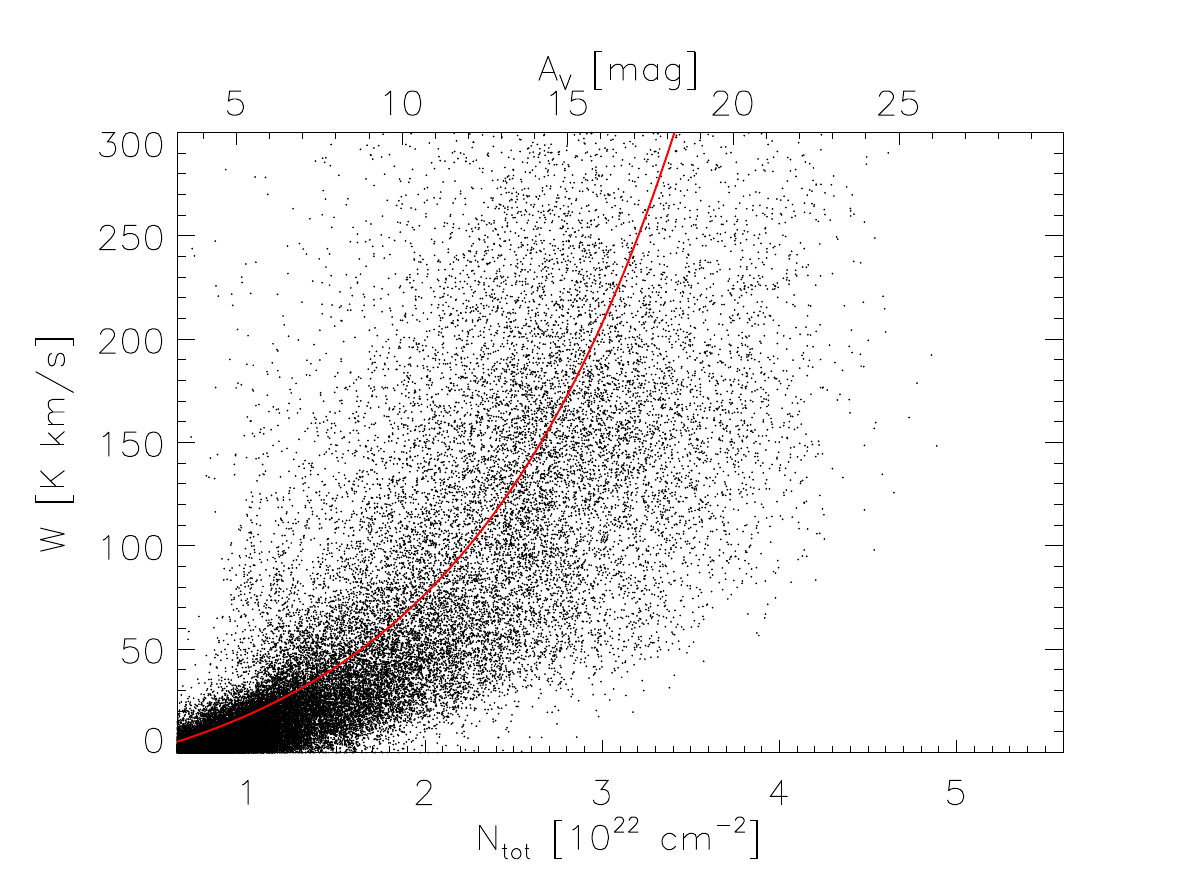}
\caption{CO intensity W as a function of the total column density $\rm N_{H}$
(bottom abscissa) or the extinction $\rm A_{V}$ (upper abscissa). The red line
indicates the median values.}
\label{akw}
\end{figure}

\begin{figure}[!htbp]
   \includegraphics[width=9cm]{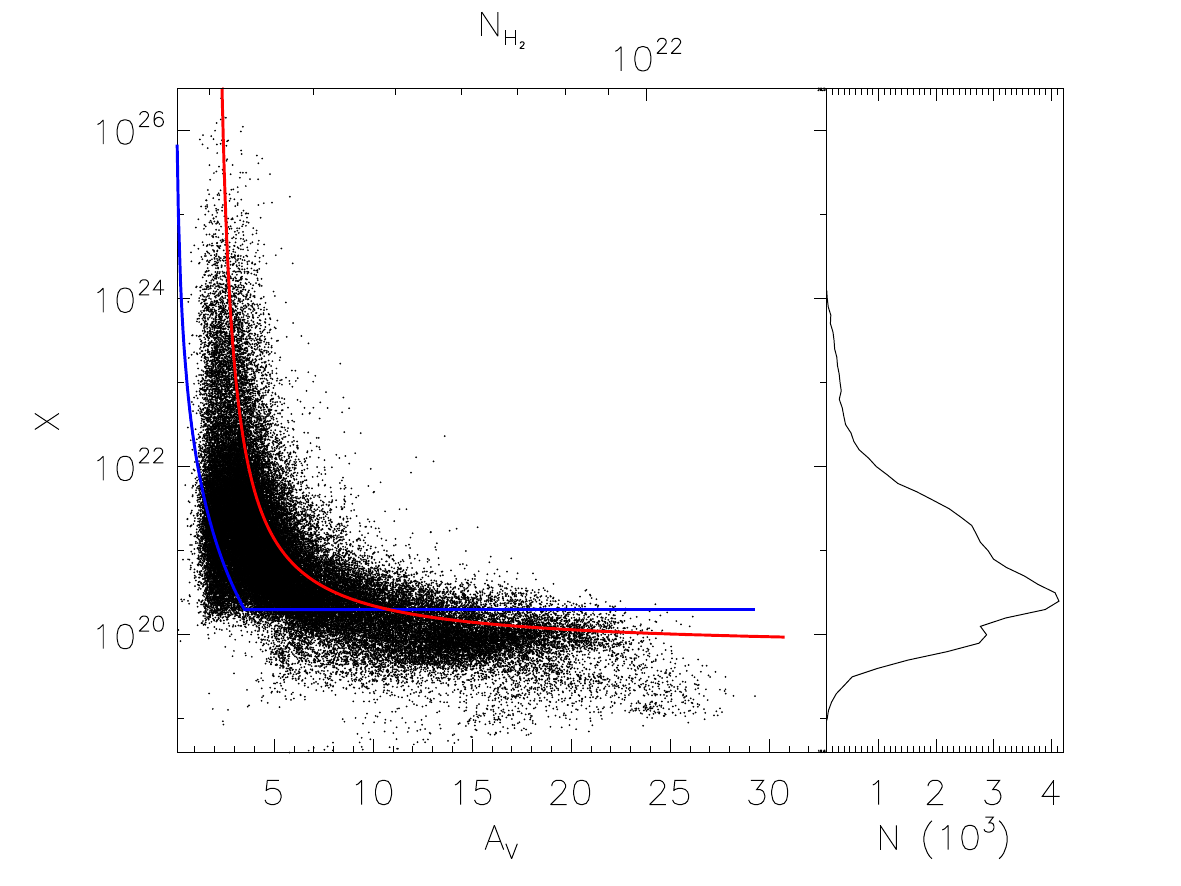}
\caption{X-factor plotted against $\rm N_{H_2}$ (left panel) and the histogram of the X-factor (right panel). Here we adopted $Z=1.0\,Z_o$. The blue line shows the relation from \citet{glover2011}.} 
\label{nh2x}
\end{figure}

\begin{figure}[!htbp]
   \includegraphics[width=9cm]{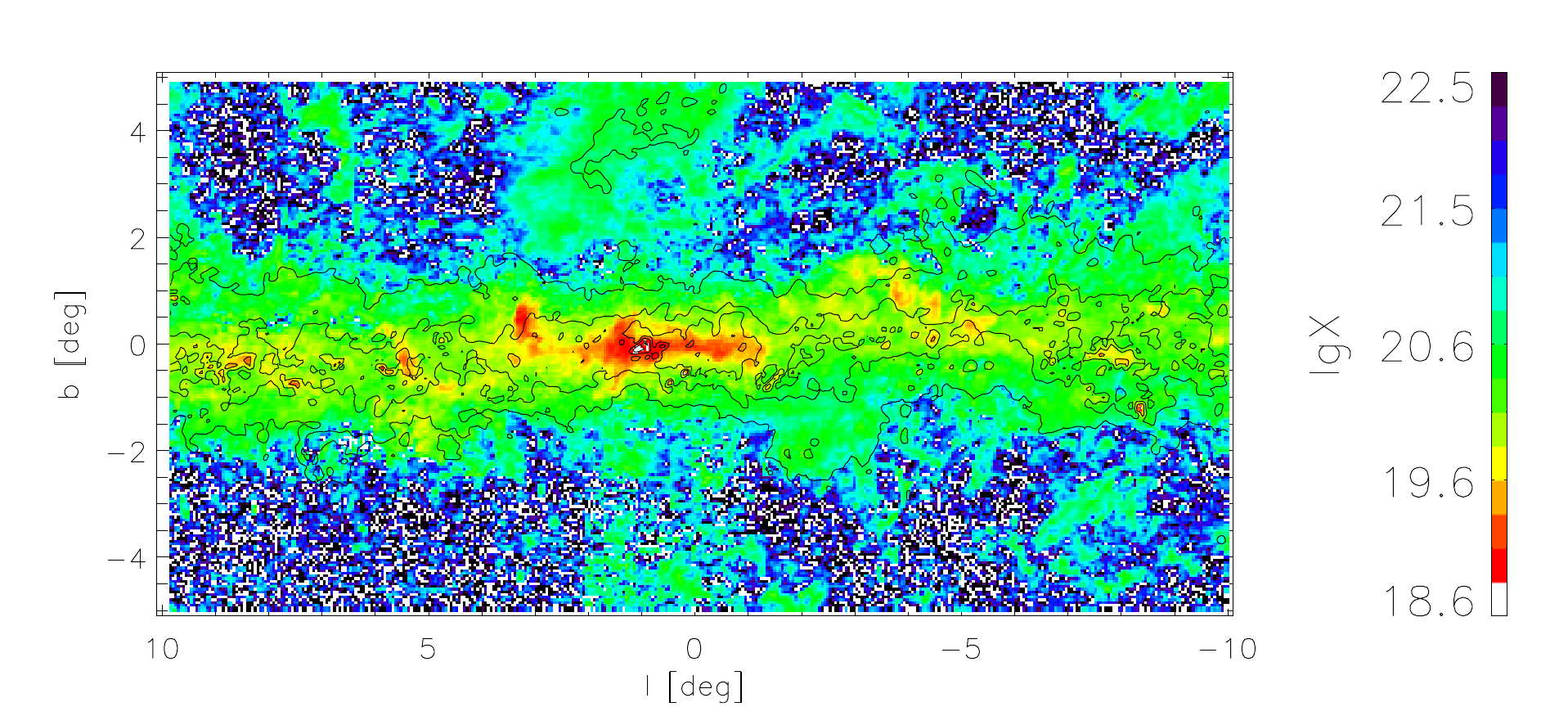}
    \includegraphics[width=9cm]{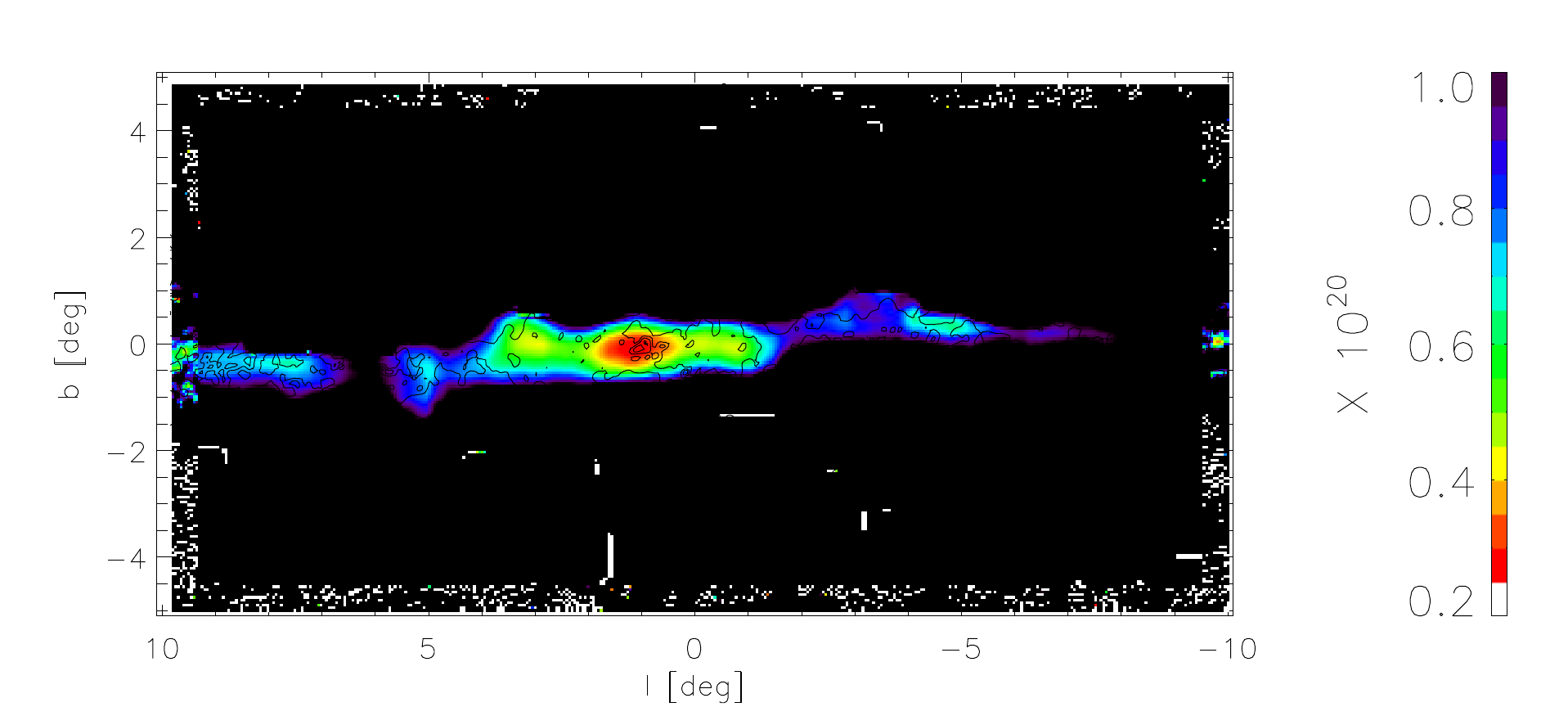}
\caption{Top panel: The variation of the X-factor as a function of Galactic longitude and latitude. The colour scale is logarithmic. Lower panel:  smoothed map to a resolution of 40'\,$\times$\,40' with a linear colour scale similar as in \citet{regan2000}.}.
\label{nh2xa}
\end{figure}

%Recently, \citet{enokiya2013} discovered newly detected molecular filaments (MFs)  using high sensitive $\rm ^{12}CO$  observations with NANTEN2 associated with the infrared Double Helix Nebula (DHN) towards the Galactic Centre. We use this data set to compare it with our dust extinction map.
The observations were made in the OTF mode and the final r.m.s noise fluctuations
are 0.5--1.0\,K for a velocity resolution of $\rm 0.65\,km\,s^{-1}$. The data cover the inner region of the Bulge
($\rm |l| < 10, |b| < 5$). We refer here to \citet{enokiya2013} for a more detailed description of the data set.

We used  the integrated CO map, which is in units of the brightness temperature (K). To compare our extinction map with the CO data, we smoothed the CO data  to the same resolution as ours.  Figure~\ref{akco} shows the comparison of the dust extinction (lower panel) and the CO data (right panel). The gas and dust distribution concentrated
on the Galactic plane ($\rm |b| < 0.5$) are in general similar. The highest CO intensity and the highest the dust extinction occurs around the central molecular zone.
% Interestingly, we see  also regions with strong  CO emssion but lower  $\rm A_{Ks}$ extinction, e.g. at $\rm l = 1.5^{o}, b= 0.0^{o}$ and vice versa, i.e.: dust extinction  with hardly any CO emission (e.g. at $\rm -2 < l < -1^{o}, b= 0.0^{o}$)

Extinction measurements provide estimates of the total amount of dust for different lines of sight. Assuming a certain reddening law  (and an assumption for the dust-to-gas ratio), the total gaseous column follows directly from the amount of extinction $\rm A_{V}$ (\citealt{Bohlin1978}). 

% Furthermore, if the temperature and the metallicity abundance is assumed, we can use the extinction map to compare with the CO map. \citet{enokiya2013} obtained very high resolution CO emission maps for the Galaxy centre using NANTEN. We presented in Fig.~\ref{akco} the maps for both the integrated CO map and our extinction map for the area of Galaxy centre. The integrated CO map is in the unit of brightness temperature \,K and we here shows the logarithmic of W for easier comparsion. As \citet{enokiya2013} have very high resolution, the CO map has been smoothed and the extinction map have been interpolated to the same resolution (3' $\times$ 3'). From Fig.~\ref{akco} we can see that the features of the gas and the dust, which have similar distribution.  

% Figure.~\ref{akco} shows the CO map together with the extinction map for the inner bulge toward the galactic centre. Fig.~\ref{akco2} shows Correlation between the extinction and the CO dust. The black line is the best linear fit.
% \begin{equation}
%  lgCO = 0.564941  +   0.930750 * A_{Ks},
% \end{equation}

In most of the molecular clouds  almost  all the hydrogen is nearly  molecular and can therefore  be expressed as the total column density of the hydrogen nuclei $N_{H}$. We here applied the simple conversion between $N_H$ and $A_V$ (\citealt{shetty2011a}),
\begin{equation}
 N_H=A_V*1.87 \times 10^{21} /(\frac{Z}{Z_{\odot}}),
\end{equation}

where Z is the metallicity of the gas.

To convert $A_{Ks}$ into $A_V$ we used the extinction law $A_V=A_{Ks}/0.089$ from \citet{glass1999}. The metallicity in the Galactic bulge is still debated.
 Recent measurements indicate a solar iron abundance, but
a high $\alpha$-element value (see e.g. \citealt{zoccali2008}; \citealt{hill2011}; \citealt{gonzalez2011}; \citealt{bensby2013}). We adopted  $\rm Z=1.0 \times Z_{\odot}$. 
Fig.~\ref{akw} shows the integrated CO intensity W as a function of the total column density  $\rm N_H$ (bottom abscissa) or the extinction $A_V$. The red line indicates the median values.  W  increases with the increasing total column density. It shows a similar trend as predicted by detailed radiative transfer calculations of molecular clouds (see Fig. 1  from \citealt{shetty2011a}).
We also note an increased dispersion in W with increasing extinction. A part of the high dispersion might  also be due to the high metallicity dispersion in the Galactic bulge ($>$ 0.5\,dex). In addition, the gas pressure and  temperature are higher, especially around the Galactic centre region, which  leads to much higher CO intensities.

%. It shows the same trend as the model from \citet{shetty2011a}. But in contrast of their high-density ran (n1000), we obtain a much high increasing rate for the W. This is not only because of the high metallicity we adopted (1.7 compared their 1.0 Z0). One important reason could be that for \citet{shetty2011a}, they have the problem of the saturation of CO emission while in our case, the CO can get very high values commonly. In contrast, we can see the limit of the extinction for our case. The  saturation at high extinction makes the tail of the red line goes to very cliff.

%First, there is a general trend of increasing intensity with increasing column density, though the slopes differ between the various simulations. For the high-density run (n1000) shown in Fig. 1(c), the CO intensities do not demonstrate any clear trend with increasing AV. Rather, the vast majority of the intensities reach a threshold value of ~65 K km s~1. This saturation of CO intensities is expected to occur at high densities since the CO line becomes optically thick. Saturation is also found at the highest extinctions in the Milky Way simulation (n300, Fig. 1a). Indeed, saturation at high CO intensities has been observed in MCs in the solar neighbourhood 

To use CO as a proxy for $\rm H_{2}$ we used the  X-factor, which
is approximately constant for the Galactic disk molecular clouds. However, 
lower densities and metallicities
increase the X-factor to higher values because of  the lower CO excitation
and CO/$H_{2}$ ratios.

To obtain the X-factor, we derived the column density of $\rm H_2$.
 We used   Tab. 2 from \citet{glover2011} to convert  $N_{H}$ into  $N_{H_2}$. For the high densities in the Galactic bulge, we chose the model n1000, which means
 \begin{equation}
 N_{H_2}=N_{H}*0.998/2.
\end{equation}
Using the definition from Eq.~1, the X-factor can be calculated as follows:
 \begin{equation}
 X=N_{H_2}/W=(A_{Ks}/0.089) \times 0.935 \times 10^{21} / W.
\end{equation}
We obtain an average value for X=$\rm 2.5 \pm 0.47 \times 10 ^{20}cm^{-2}K^{-1}km^{-1}s$. This value is consistent with the  canonical value of 
 X=$\rm 2 \times 10 ^{20}cm^{-2}K^{-1}km^{-1}s$  determined observationally for the Milky Way (\citealt{dame2001}).

 Figure~\ref{nh2x} shows how the X-factor changes with extinction.The red line shows the following relation

\begin{equation}
 X=10^{20.0} - 10^{100*A_V + 150}.
\end{equation}

 \citet{glover2011} have demonstrated that the CO-to-$\rm H_{2}$ conversion factor is  determined primarily by the mean extinction of the cloud and that it is almost constant for $\rm A_{V} > 3$. They reported that the X-factor declines from X=$\rm 10 ^{24}$ to X=$\rm 3 \times 10 ^{20}cm^{-2}K^{-1}km^{-1}s$ for $\rm A_{V}$s in the range of $\rm 0.2 < A_{V} < 3$. We note  a steady decrease
of the X-factor with increasing extinction over the  $A_{V}$-range $\rm A_{V} < 10$ while  a kind of flattening occurs at $\rm A_{V} > 10$. 

 Our results show a very high dispersion of the X-factor for a given $\rm A_{V}$ that leads to a broad range of the X-factor  of nearly a  factor of 100 because of the  extreme conditions in the Galactic bulge  with its well-known high metallicity and velocity dispersion (e.g. \citealt{gonzalez2013}; \citealt{howard2008}; \citealt{babusiaux2010}). The X-factor is known to depend on   gas  properties such as the temperature, density, metallicity, and its velocity field  (\citealt{glover2011}; \citealt{papadopoulos2012}). \citet{narayanan2013} used a combination of high-resolution galaxy evolution simulations to resolve
giant molecular clouds (GMCs) and perform 3D molecular line radiative transfer calculations and  found that the X-factor remains constant as a function of Galactocentric radius (see their Fig.~2). Figure~\ref{nh2xa} shows the variation of the X-factor as a function of the Galactic longitude and latitude. The very strong variation along individual lines of sight in the upper panel is expected  because of  large differences in the small-scale molecular cloud properties. The lower panel shows the map smoothed to a resolution of 40\arcmin $\times$ 40\arcmin, which corresponds to $\sim 100$\,pc, a minimum spatial scale typically used in extragalactic studies where the X-factor is applied to an ensemble of molecular clouds (\citealt{regan2000}; \citealt{papadopoulos2012}). Even in the smoothed map we clearly see the decrease of the X-factor in the Galactic centre region which indicates
 that the X-factor  varies across the entire Bulge. The dispersion in X-factor measured from our maps is  higher  by a factor of 2 than that reported by  \citet{narayanan2013}.

It is interesting to compare our smoothed X-factor map (lower panel in Fig. \ref{nh2xa}) with the maps of the central regions of nearby spiral galaxies (\citealt{regan2000}). In these maps the X-factor  is much lower (5–20 times) in the central regions  than the value used locally in the Milky Way. 
   The low X-factor values in the central regions of spiral galaxies agrees with our results, while the  extremly low X-factors we derive for the central molecular zone in the Milky Way is typically not observed, which indicates extreme conditions in the central molecular zone.

%We think that 
%metallicity could be responsible for  this additional dispersion.

%Metallicity maps such as 
%those from \citet{Gonzalez2013} would be very useful to get a better estimate of the X-factor.

%The decreasing of X with the increasing extinction or $N_{H_2}$ is found for our result. This results fits the low densities or metallicities of \citet{shetty2011a}, but opposite for their high density model (n1000), which was due to their line saturation \citep{shetty2011a}. \citet{glover2011}'s result shows the same trend as ours. But we do not see the trend to get a constant value due to the extremely environment of the Galaxy centre. We can see that the X-factor goes a relatively small range for the Galaxy centre from the histogram of the X (the right panel of Fig.~\ref{nh2x}).

\section{Summary}
Using an improved version of the Besan\c{c}on model, we presented  high-resolution 3D extinction maps (6\arcmin $\times$ 6\arcmin)  for the entire VVV bulge region. All extinction maps are available  online at the  CDS (Table 1) or  through the BEAM calculator webpage (http://mill.astro.puc.cl/BEAM/calculator.php). Owing to the high sensitivity of the VVV data, we were able to trace high extinction until 10\,kpc. Our maps integrated along the line of sight up to 8\,kpc  agree excellently well  with the 2D maps from \citet{schultheis1999}, \citet{dutra2003}, \citet{nidever2012}, and \citet{gonzalez2012}. These maps show the same dust features and consistent $A_{Ks}$ values. These 3D maps are  a powerful tool in combination with  stellar population synthesis models such as TRILEGAL.

Using high-resolution $^{12}CO$ maps from NANTEN2 (\citealt{enokiya2013}) in the central molecular zone, we detected similar  features in the gas and the dust. Assuming that all hydrogen is molecular and
 the mean metallicity is around solar, we were able to determine the X-factor with a mean value of X=$\rm 2.5 \pm 0.47 \times 10 ^{20}cm^{-2}K^{-1}km^{-1}s$.  Our mean value is consistent with the canonical value of molecular clouds in the Milky Way (\citealt{dame2001}) and a high dispersion around the mean.

%Contrary to \citet{glover2011}, we see a decrease of the X-factor with increasing extinction.
%than the previous work because of the high density and high metallicity abundance of the Galaxy Cente. We also showed the variation of X with the extinction.

\begin{acknowledgements}
We  thank the anonymous referee for his/her helpful comments.
We thank Padelis Papadopoulos for the very informative discussion about the molecular gas and X-factor measurements in galaxies.
We gratefully acknowledge use of data from the ESO Public Survey program ID 179.B-2002 taken with the VISTA telescope, data products from the Cambridge Astronomical Survey Unit, the BASAL CATA Center for Astrophysics and Associated Technologies PFB-06, the MILENIO Milky Way Millennium Nucleus from the Ministry of Economy’s ICM grant P07-021-F, and the FONDECYT the Proyecto FONDECYT Regular No. 1130196. B.Q.C was supported by a scholarship of the China Scholarship Council (CSC).
\end{acknowledgements}

\bibliographystyle{aa}
\bibliography{hr3dextin_revised_accepted}

\end{document}